\documentclass[iop,apj,tighten]{emulateapj}
\usepackage[breaklinks,colorlinks,urlcolor=blue,citecolor=blue,linkcolor=blue]{hyperref}
\usepackage{cleveref}
\usepackage{graphicx}
\usepackage{bm}
\usepackage{amsmath,amssymb}
\usepackage{appendix}
\usepackage{color}
\newcommand{\Msun}{{\rm M}_\odot}

\shorttitle{Binary evolution in Active Galactic Nuclei}
\shortauthors{Tagawa et al.}

\begin{document}
\title{Spin Evolution of Stellar-mass Black Hole Binaries in Active Galactic Nuclei}

\author{Hiromichi Tagawa\altaffilmark{1}, Zolt{\'a}n Haiman\altaffilmark{2}, Imre Bartos\altaffilmark{3}, Bence Kocsis\altaffilmark{1}}
\affil{\altaffilmark{1}Institute of Physics, E{\"o}tv{\"o}s University, P{\'a}zm{\'a}ny P.s., Budapest, 1117, Hungary\\
\altaffilmark{2}Department of Astronomy, Columbia University, 550 W. 120th St., New York, NY, 10027, USA\\
\altaffilmark{3}{Department of Physics, University of Florida, PO Box 118440, Gainesville, FL 32611, USA}
}
\email{E-mail: htagawa@caesar.elte.hu}

\begin{abstract} 
The astrophysical origin of gravitational wave (GW) events is one of the most timely problems in the wake of the LIGO/Virgo discoveries. In active galactic nuclei (AGN), binaries form and evolve efficiently by dynamical interactions and gaseous dissipation. Previous studies have suggested that binary black hole (BBH) mergers in AGN disks can contribute significantly to BBH mergers observed by GW interferometers. Here we examine the distribution
of the effective spin parameter $\chi_\mathrm{eff}$ of this GW source population.
We extend our semi-analytical model of binary formation and evolution in AGN disks by following the evolution of the binary orbital angular momenta and black hole (BH) spins. BH spins change due to gas accretion and BH mergers, while the binary orbital angular momenta evolve due to gas accretion and binary-single interactions. We find that the distribution of 
$\chi_\mathrm{eff}$ predicted by our AGN model is similar to the distribution observed during LIGO/Virgo O1 and O2. 
On the other hand, 
if radial migration of BHs is inefficient, 
$\chi_\mathrm{eff}$ is skewed toward higher values compared with the observed distribution,
because of the paucity of scattering events that would randomize spin directions relative to the orbital plane.
We suggest that high binary masses and the positive correlation between binary mass and the standard deviation 
of $\chi_\mathrm{eff}$ for chirp masses up to $\approx 20\,\Msun$, 
can be possible signatures for mergers originating in AGN disks. 
Finally, hierarchical mergers in AGN disks
naturally produce properties of the recent GW event GW190412, including a low mass ratio, a high primary BH spin, and a significant spin component in the orbital plane. 
\end{abstract}
\keywords{
binaries: close
-- gravitational waves 
--galaxies: active
-- methods: numerical 
-- stars: black holes 
}

\section{Introduction}

Recent detections of gravitational waves (GWs) have shown evidence for a high rate of black hole (BH)-BH and neutron star (NS)-NS mergers in the Universe
\citep{TheLIGO18,Venumadhav19}. 
However, the proposed astrophysical pathways to mergers remain highly
debated.  Indeed there are currently an exceedingly large number of
such possible pathways, with widely different environments and
physical processes.  A comprehensive list of these currently includes 
isolated binary evolution \citep[e.g.][]{Dominik12,Kinugawa14,Belczynski16,Spera19} 
accompanied by mass transfer \citep[][]{Pavlovskii17,Inayoshi17,vandenHeuvel17}, common envelope ejection \citep[e.g.][]{Paczynski76,Ivanova13}, envelope expansion \citep{Tagawa18}, chemical homogeneous evolution in a tidally distorted binary  \citep{deMink16,Mandel16,Marchant16}, 
evolution of triple or quadruple systems
\citep[e.g.][]{Silsbee17,Antonini17}, 
gravitational capture 
\citep[e.g.][]{OLeary09,Gondan18a,Rasskazov19}, 
dynamical evolution in open clusters
\citep[e.g.][]{Banerjee17,Kumamoto18} 
and dense star clusters 
\citep[e.g.][]{PortegiesZwart00,Samsing14,OLeary16,Rodriguez16,DiCarlo19}, 
and dynamical interaction in gas-rich nuclear region
\citep[e.g.][]{Bartos17,Stone17,McKernan17,Tagawa19}.

Galactic nuclei are the densest environments of stars and compact
objects in the Universe \citep[see][for a recent review]{Neumayer20}. 
In an active galactic nucleus (AGN), a high-density gas disk forms within 0.1--10 pc \citep[][]{Burtscher13} around a central super-massive BH (SMBH).  
Several authors have recently pointed
out that these environments are conducive to forming compact-object
binaries.  This ``AGN disk channel'' has received increasing
attention in the wake of the LIGO/Virgo discoveries, as a possible explanation
for some of the LIGO/Virgo events.  
In particular, \citet{McKernan12,McKernan14} predicted the formation of intermediate-mass BHs in AGN disks due to collisions of compact objects.
\citet{Bartos17} have proposed a pathway for binary BH (BBH) mergers in AGN disks in which binaries are captured by an accretion disk within $\sim0.01$ pc from the SMBH due to linear momentum exchange during disk-crossing, and after that, binaries are hardened by gas dynamical friction by an AGN disk and type I/II torques by circumbinary disks. 
\citet{Stone17} have proposed another pathway, in which in-situ formed binaries at $\sim$pc scale evolve via binary-single interactions with a disk stellar component and type I/II torques from  circumbinary disks. 
\citet{Leigh18} showed that fewer than ten binary-single interactions
are sufficient to drive hard binaries with a binary separation of
$s\lesssim10$ AU to merger. 
\citet{McKernan20} investigated the evolution of neutron stars and white dwarfs in addition to BHs, and estimated the rate of mergers among these objects. 
\citet{Ford19} suggested that the AGN models can be constrained from GW observations. 
\citet{Bellovary16} suggested that BHs accumulate and merge with each other in migration traps at $20-300$ Schwarzschild radii from the central SMBH, where the sign of the torque from the AGN disk changes. 
\citet{Secunda18}, \citet{Yang19a,Yang19b} and \citet{Gayathri19} investigated the properties of mergers in migration traps. 
\citet{Tagawa19} investigated how binaries form and merge in AGN disks by performing self-consistent one-dimensional $N$-body simulations combined with semi-analytical prescriptions of the relevant processes. They found that binaries form efficiently in the inner regions ($\lesssim$ pc; but well outside the migration traps) of AGN disks, due to the dissipation of relative velocities of unbound pairs of BHs via gas drag (``gas-capture'' binaries), 
and merge at $\sim 0.01$ pc from the SMBH, where gaps form around BHs and so interactions among compact objects become frequent, similarly to migration traps.

Motivated by the above, in the present paper, we investigate whether
the AGN disk channel can be distinguished from other formation
pathways.  Previous work proposed distinguishing features based on
spatial associations with bright AGN \citep{Bartos17NatCom,Corley19}, large chirp masses \citep{Tagawa19}, redshift evolution \citep{Yang20}, acceleration of the binary's center of mass \citep{Meiron17,Inayoshi17b,Wong19}, or 
gravitational lensing \citep{Kocsis2013,DOrazio19}. 
One feature that has not yet been studied in detail in this
channel is the expected distribution of BH spins.  The effective spin
parameter $\chi_\mathrm{eff}$ (which is the sum of the projection of
binary spins onto the binary orbital angular momentum) has been shown
to provide useful information to constrain other compact-object merger
pathways~\citep{Stevenson17,Vitale17,Talbot17}.

The $\chi_\mathrm{eff}$ distribution inferred from the observed GW events 
prefers low values~\citep{TheLIGO18}, 
which suggests low natal BH spins or 
random directions between the binary orbital angular momentum and the BH spins \citep{Farr17}. 
On the other hand, several events are reported to have high or low $\chi_\mathrm{eff}$ values \citep[][]{Zackay19,Zackay19b,LIGO20_GW190412}. 
\citet{Safarzadeh20} suggested that there
is a negative and positive correlation between mass and
the mean and the dispersion of $\chi_\mathrm{eff}$, respectively. 
For the evolution of isolated binaries, 
the low observed $\chi_\mathrm{eff}$ values may be reproduced if 
the angular momentum transport within the stars is highly efficient \citep{Qin18,Bavera19}. 
In globular clusters, the orbital angular momentum directions of binaries are randomized by binary-single interactions, which predicts 
a $\chi_\mathrm{eff}$ distribution symmetric around $\chi_\mathrm{eff}=0$ and favoring low values
\citep{Rodriguez18PRL,ArcaSedda18}. 
For triple systems, the Kozai mechanism can cause BH spin misalignment \citep{Liu17,Liu18,Liu19,FragioneKocsis19b}.

Several recent studies have investigated the properties (e.g. $\chi_\mathrm{eff}$, mass ratio, binary mass) of binary mergers in AGN disks. 
For mergers in migration traps, 
\citet{Yang19a,Yang19b} and \citet{McKernan19} performed Monte Carlo simulations, and predicted that typical mergers have a low mass ratio of $q\sim0.2$ and significant effective spin $|\chi_\mathrm{eff}|\sim 0.4$. 
\citet{Gayathri19} showed that mergers in migration traps can explain the values of $\chi_\mathrm{eff}$ and $m_\mathrm{chirp}$ for GW170817A. 
\citet{Secunda18,Secunda20}  investigated the dynamical evolution of compact objects around migration traps by directly following their orbits in numerical simulations.

Focusing on mergers occuring outside 
migration traps, 
\citet{McKernan19} estimated the mass and 
spin distribution for mergers among binaries formed during close encounters in AGN disks, by assuming that binaries are always aligned or anti-aligned with the AGN disk, and BH mergers are much faster than 
the growth of BH spins by gas accretion. 
They found that the distribution of $\chi_\mathrm{eff}$ is symmetric around zero, and the dispersion of $\chi_\mathrm{eff}$ is determined by the magnitude of initial BH spins. 
They predicted a variety of the merged mass distribution depending on adopted initial mass functions, which are roughly consistent with results by \citet[][hereafter Paper~I]{Tagawa19} 
However, in Paper~I, BH mergers were found to be less rapid, as gas dynamical friction becomes inefficient as the binary separation shrinks, and because Paper~I employed an updated version of type I/II migration theory,which predicts weaker torques compared with previous ones \citep[e.g.][]{Duffell14,Kanagawa18}. 
Also, Paper~I 
suggested that the orbital angular momenta of the binaries at mergers are often misaligned with the AGN disk 
due to frequent hard binary-single interactions. 
Since the merger timescale of BBHs is 
comparable to the timescale of gas accretion onto BHs, 
the evolution of the BH spins needs to be
explicitly followed, accounting for both effects. 
Furthermore, to determine the $\chi_\mathrm{eff}$ distribution for mergers in AGN disks, it is 
necessary to also follow 
the orbital angular momenta 
of the binaries, again taking into account both binary-single interactions and gas accretion.

In this paper, we determine the distribution of $\chi_\mathrm{eff}$ for binary mergers in AGN disks,  by incorporating the evolution of BH spins and the binary orbital angular momenta into the semi-analytical prescriptions and one-dimensional N-body simulations used in Paper~I.  We find that the frequency of binary-single interactions, 
the angular momentum of the captured gas, and the initial BH spin directions strongly influence the $\chi_\mathrm{eff}$ distribution. 
The rest of this paper is organized as follows. 
In $\S$~2, we describe the numerical scheme and the setup of the simulations. 
We present our main results in $\S$~3, and summarize our conclusions in $\S$~4.

\section{Method}

To derive the $\chi_\mathrm{eff}$ distribution at mergers, 
the evolution of the dimensionless BH spins (${\bm a}$) and the binary orbital angular momentum directions (${\hat{\bm J}}_\mathrm{bin}$)\footnote{{we use the usual notation for unit vectors $\hat{\bm x} = {\bm x}/|\bm x|$}} 
needs to be followed
since $\chi_\mathrm{eff}$ is the sum of the projection of mass-weighted binary spins onto the binary orbital angular momentum, 
\begin{equation}
\chi_\mathrm{eff}=\frac{m_1{\bm a}_1+m_2{\bm a}_2}{m_1+m_2}\cdot \hat{\bm J}_\mathrm{bin}.
\end{equation}
Here $m_1$ and $m_2$ are the masses
and ${\bm a}_1$ and ${\bm a}_2$ are the spins of
the binary components. To model the evolution of the BH spins and the binary orbital angular momenta, 
we perform one-dimensional $N$-body simulations 
combined with semi-analytic prescriptions.
In the following sections, we first give a brief overview of our model and then describe its ingredients in more detail.

\subsection{Overview of model}

In this section, we summarize our model, whose details are presented in Paper~I. 
We consider a system describing a galactic nucleus, consisting of the
following five components: (1) a central SMBH, (2) a gaseous accretion disk around the SMBH (``AGN disk''), (3) a spherical stellar cluster, (4) a flattened
cluster of BHs, and (5) stars and BHs inside the AGN disk, referred to
as the ``disk stellar'' and ``disk BH'' components. 
To follow the time-evolution of the BHs in this system, focusing on
their capture by the disk, and the formation, evolution, and disruption of BH binaries in the disk, we run one-dimensional $N$-body simulations combined with a semi-analytical method. 
We introduce $N$-body particles representing 
either single objects or binaries, and for each particle, we follow 
its radial position from
the central SMBH, as well as its radial velocity, together with
the evolution of the binaries' separation. 
The other two 
spatial directions are followed only statistically. 
In this paper, we additionally follow the evolution of BH spins ($\S$~\ref{sec:bh_spin}) and the orbital angular momentum directions of binaries ($\S$~\ref{sec:am_ev}).

For the AGN disk, we employ the model proposed by \citet{Thompson05}, as adopted in the earlier work by \citet{Stone17}. This represents a Shakura-Sunyaev $\alpha$-disk with a constant viscosity parameter
$\alpha$ and accretion rate in the region where it is not
self-gravitating.  The model describes a radiatively efficient, geometrically thin, and optically thick disk and
extends the disk to pc scales with a
constant Toomre parameter in the self-gravitating regime, assuming that it is heated and stabilized by radiation pressure and supernovae from in-situ star formation. We assume that stars and BHs form in the disk at the rate required to stabilize the AGN disk, and some fraction of BHs are initially formed in binaries (see parameter settings in Table~\ref{table_parameter_model}). 

We assume that the AGN disk is surrounded by a spherically symmetric star cluster, with a total mass  $\approx 3$ times that of the central SMBH within $\sim 3\,\mathrm{pc}$, and with a density profile matching those of a nuclear cluster observed in the Galactic center.
We further include a flattened BH cluster component, 
which has a steeper density profile and a smaller velocity dispersion compared with those of a spherically symmetric star cluster due to mass segregation \citep[e.g.][]{Hopman06,Szolgyen18}.

Our model tracks the properties of the BH population, including physical processes due both to the presence
of gas and to multi-body dynamical interactions, as follows.

For the interaction with gas, the velocities of all BHs relative to the local AGN disk decrease due to accretion torque and to gas dynamical friction.
For binaries of stellar-mass BHs, the binary separation evolves due to gas dynamical friction from the AGN disk and to type I/II migration torque from a small circumbinary disk that forms within the Hill sphere of the binary. Binaries efficiently form in the disk due to gas dynamical friction during two-body encounters (a process we dubbed ``gas-capture binary formation''). 
The radial positions of BHs are also allowed to evolve due to type I/II torques from the AGN disk. Gas accretion affects BH spins and the orbital angular momentum directions of binaries according to newly added prescriptions ($\S$~\ref{sec:spin_gas_accretion} and $\S$~\ref{sec:gas_acc_am}, respectively).

We also account for dynamical interactions with single stars and BHs and BH binaries.  The binaries' separations and velocities evolve due to binary-single interactions, and the velocities of all BHs additionally evolve due to scattering. 
The evolution of the orbital angular momentum directions of binaries during binary-single interactions are additionally followed with newly added prescriptions ($\S$~\ref{sec:bs_am}).
Binaries form due to three-body encounters, and are disrupted by soft binary-single interactions. 
We also account for GW emission from binaries, which reduces their
separation rapidly once they are sufficiently tight. 
For simplicity, the eccentricity evolution is ignored and orbits around
the SMBH and binary orbits are both assumed to be circular.

The interested reader is encouraged to consult Paper~I for detailed descriptions of the above model and its ingredients. In the following sections, we only describe the new prescriptions which we added to Paper~I, in order to follow the evolution of $\chi_\mathrm{eff}$ of each BH binary.

\subsection{BH spin evolution}
\label{sec:bh_spin}

BH spin is characterized by the dimensionless spin parameter 
${\bm a}=c{\bm J}_\mathrm{BH}/Gm_\mathrm{BH}^2$, where $G$ is the gravitational constant, $c$ is the speed of light, $m_\mathrm{BH}$ is the
mass and ${\bm J}_\mathrm{BH}$ is the angular momentum of the BH. 
In this section, we describe the initial distribution and the subsequent evolution of BH spins.

\subsubsection{Initial BH spin distribution}

In our model, there are two types of BHs differentiated by their origin:
BHs formed before the beginning of the current AGN phase (pre-existing BHs), and BHs formed during the current AGN phase (in-situ formed BHs). 
Pre-existing BHs are distributed in nuclear star clusters, 
but have a density profile that is steeper \citep[e.g.][]{Hopman06,Freitag06,Keshet09} and velocity dispersion that is smaller \citep{Szolgyen18} compared to those of typical-mass stars. Pre-existing BHs are expected to have random spin directions since they presumably formed by the disruption of globular clusters \citep[e.g.][]{Mapelli16}, or by the fragmentation of previous AGN disks 
or disks in non-active phases 
whose orientations differed from the current one.

On the other hand, in-situ BHs form in the outer regions of the AGN disk, 
and could have their spins directed along the angular momentum of the AGN disk.  This may be expected if these BHs
form from, or efficiently accrete, gas whose angular momentum direction is the same as that of the background AGN disk.
This assertion might be justified by the analogy with the planets in the Solar system, all of which except for Venus and Uranus spin in the same direction as their orbital motion to within $30^{\circ}$. 

The typical values of the initial BH spins are highly uncertain. 
The progenitors of some BBHs, BHs in high-mass X-ray binaries,
are observed to have high spin (see, e.g. \citealt{Miller15} for a review). 
However, we do not have any information on the spins of isolated single BHs or for heavier BHs with masses similar to those discovered by GW observations.

We therefore consider several distributions for the initial BH spin ${\bm a}$. For the direction of the initial BH spin $\hat{\bm a}$, we examine two models: (i) the spin direction $\hat{\bm a}$ is random, (ii) $\hat{\bm a}$ is directed along the angular momentum of the AGN disk $\hat{\bm J}_\mathrm{AGN}$ (the latter defined with respect to the SMBH), where we fix $\hat{\bm J}_\mathrm{AGN}=\hat{\bm z}$, i.e. along the $z$-axis.
For the magnitude of the initial BH spin $|{\bm a}|$, 
we examine the full range of values between 0 and 0.99 (models~M1--M7; see Table~\ref{table_results} below). 
In the fiducial model (M1), 
$a_0=0$ for both pre-existing and in-situ BHs \citep[e.g.][]{Fuller19}. 
In models~M2--M7, we instead assume $a_0=0.1$--$0.99$, respectively.
In all six models~M2--M7, the spin direction of the pre-existing and in-situ formed BHs were assumed random and aligned with the AGN disk, respectively.
In models~M8 and M9, we adopt $a_0=0.7$ for all BHs \citep[e.g.][]{Shibata02}, 
but assume that 
{\em all} BH spins are random (model~M8), or  aligned with the AGN disk (model~M9).

Note that in our models the initial BH mass is below $15\,\Msun$, which may be expected for high-metallicity environments as in AGN. Unlike in other BH formation channels with heavier BHs \citep[e.g.][]{Gerosa18}, at these lower masses there is no apriori anti-correlation between mass and spin. The BH masses and spins change significantly from their initial values during the evolution in AGN due to gas accretion and mergers in our simulations.

\subsubsection{Gas accretion}
\label{sec:spin_gas_accretion}

In our model, BHs capture gas from the AGN disk while they are moving in or crossing the disk. 
Some fraction of captured gas is assumed to accrete onto BHs through circum-BH disks (for single BHs) or mini-disks (for BBHs, fed from circumbinary disks). 
During such accretion processes the binary mass, velocity, and separation, as well as the BH spins all evolve.

The spin values $a=1$ and $-1$ represent a maximally spinning BH, and the sign of $a$ is defined so that for $a>0$ the BH is spinning in the same direction as the inner accretion disk, and for $a<0$ the spin is in the opposite direction. 
The spin magnitude after an accretion episode is given by
\begin{equation}
\label{eq:spin_evolution}
    a^\mathrm{f}=\frac{1}{3}
    \frac{r_\mathrm{isco}^{1/2}}{f_\mathrm{acc}}
    \left[ 4- \left ( 3 \frac{r_\mathrm{isco}}{f_\mathrm{acc}^2} - 2 \right)^{1/2} \right]
\end{equation}
\citep{Bardeen70}, where $f_\mathrm{acc} \equiv (m_\mathrm{BH}+\Delta m_\mathrm{BH})/m_\mathrm{BH}$, 
$\Delta m_\mathrm{BH} \equiv {\dot m}_\mathrm{BH}\Delta t$ is the mass accreted during the time step $\Delta t$, 
the superscript $\mathrm{f}$ stands for the values after the episode, 
and $r_\mathrm{isco}$ is the radius of the innermost stable circular orbit (ISCO) in reduced units, defined as 
\begin{equation}
\label{eq:r_isco}
    r_\mathrm{isco}=R_\mathrm{isco}/R_\mathrm{g}=3 + Z_2 \mp \sqrt{(3 - Z_1)(3 + Z_1 +2 Z_2)}
\end{equation}
with the minus sign for $a>0$ and the plus sign for $a<0$. Here $R_\mathrm{g}=G m_\mathrm{BH} /c^2$ is the gravitational radius of the BH, and the functions $Z_1$ and $Z_2$ are given by 
\begin{equation}
    Z_1= 1 + (1-|a|^2)^{1/3}[(1+|a|)^{1/3}+(1-|a|)^{1/3}],
\end{equation}
\begin{equation}
    Z_2 = \sqrt{3|a|^2+Z_1^2}. 
\end{equation}
While Eq.~\eqref{eq:spin_evolution} gives unphysical spin values when a highly spinning BH accretes a significant amount of gas ($a^\mathrm{f}>1$ or imaginary), the torque exerted by the radiation of a thin accretion disk prevents $a^\mathrm{f}>0.998$ \citep{Thorne74}. 
Fully relativistic magnetohydrodynamics simulations suggest that the spin value does not grow beyond $0.95$ during accretion from a thick disk \citep{Gammie2004,Shapiro2005}. 
We set the upper limit of $a^\mathrm{f}$ to 0.99 in our models.

When the spin angular momentum ${\bm J}_\mathrm{BH}= {\bm a} \sqrt{G m_\mathrm{BH}^3 R_\mathrm{g}}$ of a BH is misaligned with its inner disk, 
the BH induces a Lense-Thirring precession in the misaligned disc elements, which causes the inner parts of the disk and the BH spin to align.
The transition between aligned and misaligned annuli of the disk
occurs at the so-called warp radius $R_\mathrm{warp}$. 
In each time-step $\Delta t$ in our model, ${\bm J}_\mathrm{BH}$ aligns with the 
initial spin angular momentum 
of the BH plus the angular momentum $\Delta {\bm J}_\mathrm{warp}$ of the disk within the warp radius: ${\bm J}_\mathrm{BH}\rightarrow 
{\bm J}_\mathrm{BH} + \Delta {\bm J}_\mathrm{warp}$.
For a Shakura-Sunyaev disk, the warp radius is given by
\begin{equation}
\label{eq:r_warp}
    R_\mathrm{warp}/R_\mathrm{S} =  3.6\times 10^2 |a|^{5/8} m_\mathrm{BH}^{1/8}
    f_\mathrm{Edd}^{-1/4}
    \left( \frac{\nu_2}{\nu_1}\right)
    \alpha_\mathrm{SS}^{-1/2} 
\end{equation}
\citep[e.g.][]{Volonteri07}, 
where $f_\mathrm{Edd}={\dot m}_\mathrm{BH}c^2/L_\mathrm{Edd}$ is 
the accretion rate in Eddington units (without a radiative efficiency),
$L_\mathrm{Edd}$ is the Eddington luminosity, 
$\nu_1$ is the viscosity responsible for transferring angular momentum in the accretion disk, and  $\nu_2$ is the viscosity responsible for warp propagation. 
We set $\Delta {J}_\mathrm{warp}=\Delta M_\mathrm{warp}\sqrt{G m_\mathrm{BH} R_\mathrm{warp}}$ \citep[e.g.][]{Volonteri07}, 
where $\Delta M_\mathrm{warp}$ is the warped disk mass aligned with the BH spin during $\Delta t$. 
We set $\Delta M_\mathrm{warp}=\Delta m_\mathrm{BH}$, 
and $\Delta \hat{\bm J}_\mathrm{warp}$ points in the same direction as the angular momentum of the circum-BH disk $\hat{\bm J}_\mathrm{CBHD}$. 
After alignment, the magnitude of the BH spin evolves through gas accretion via Eq.~(\ref{eq:spin_evolution}).

We note that whenever the condition
\begin{equation}
\label{eq:alignment_king}
\cos \theta_\mathrm{BH,warp} < - \frac{\Delta J_\mathrm{warp}}{2J_\mathrm{BH}}
\end{equation}
is satisfied, where $\theta_\mathrm{BH,warp}$ is the angle between ${\bm J}_\mathrm{BH}$ and $\Delta {\bm J}_\mathrm{warp}$, the accretion disk within $R_\mathrm{warp}$ becomes anti-aligned with the BH spin direction \citep{King05}.

Due to the Lense-Thirring effect, the BH spin and the circum-BH disk angular momentum $\hat{\bm J}_\mathrm{CBHD}$ align
faster than how the magnitude of the BH spin grows \citep[e.g.][]{Volonteri07}. 
In this process, there are two large uncertainties: the size of the warp radius, and the direction of $\hat{\bm J}_\mathrm{CBHD}$ 
The size of the warp radius strongly depends on the ratio $\nu_2/\nu_1$ (Eq.~\ref{eq:r_warp}). 
Many studies adopted $\nu_2/\nu_1=2(1+7\alpha_\mathrm{SS})/(4+\alpha_\mathrm{SS}^2)/\alpha_\mathrm{SS}^2 \sim 85$, motivated by
analyses of low-amplitude warps \citep{Ogilvie99}. 
On the other hand, \citet{Lodato13} have shown that when considering large-amplitude warps, $\nu_2/\nu_1$ is between $\sim 2-50$ depending on $\alpha_\mathrm{SS}$ and the misalignment angle $\theta_\mathrm{BH,warp}$. 
We set $\nu_2/\nu_1$ to be a free parameter with a fiducial value of 10, and vary it from 2 to 50 (models~M10~and~M11). 

For a single BH, $\hat{\bm J}_\mathrm{CBHD}$ aligns with $\hat{\bm J}_\mathrm{AGN}$ (see results in \citealt{Lubow99} in the context of protoplanetary disks). 
On the other hand, when a BH is in a binary, $\hat{\bm J}_\mathrm{CBHD}$ aligns with the orbital angular momentum direction of the binary $\hat{\bm J}_\mathrm{bin}$ 
on the disk's viscous timescale  \citep[e.g.][]{Moody19}. 
Assuming a Shakura-Sunyaev disk, the viscous timescale is given by
\begin{align}
\label{eq:t_visc}
    t_\mathrm{vis}\sim& \frac{s^2}{\nu}\nonumber\\
    \sim& 10^2 \mathrm{yr} \left(\frac{\alpha_\mathrm{SS}}{0.1}\right)^{-4/5}
    \left(\frac{{\dot m}_\mathrm{bin}}{{\dot m}_\mathrm{Edd,bin}}\right)^{-3/10}\nonumber\\
   & \left(\frac{m_\mathrm{bin}}{20\,\Msun}\right)^{1/4}
    \left(\frac{s}{\mathrm{AU}}\right)^{5/4},
\end{align} 
\citep[e.g.][]{Frank02}, where $s$ is the binary separation, $m_\mathrm{bin}$ is the binary mass, ${\dot m}_\mathrm{bin}$ is the accretion rate onto the binary, 
${\dot m}_\mathrm{Edd,bin}=L_\mathrm{Edd}/(\eta_\mathrm{c}c^2)$ 
is the Eddington accretion rate for the binary, and $\eta_\mathrm{c}$ is the radiative efficiency. 
Due to the short viscous timescale, in our fiducial setting we assume that $\hat{\bm J}_\mathrm{CBHD}$ is the same as $\hat{\bm J}_\mathrm{AGN}$ when a BH is single, and is the same as $\hat{\bm J}_\mathrm{bin}$ when a BH is in a binary.
For comparison,
we also investigate the alternative assumption that the direction of $\hat{\bm J}_\mathrm{CBHD}$ is always aligned with $\hat{\bm J}_\mathrm{AGN}$ (model~M12).

We set the BH accretion rate 
to the minimum of the Eddington accretion rate and the Bondi-Hoyle-Lyttleton rate (Eq.~24 in Paper~I) taking into account a reduction due to the shearing motion of the nearby disk gas (Eqs.~29--32 in Paper~I). When BHs are in binaries, we 
apportion the total accretion between primary and secondary BHs 
following \citet{Duffell19}, which is updated from Paper~I 
in which we used earlier results from \citet{Farris14}.

\subsubsection{Mergers}

Following BBH mergers, the dimensionless spin parameter ${\bm a}_\mathrm{f}$ of the remnant BH depends on the spins ${\bm a}_1$ and ${\bm a}_2$ and the dimensionless orbital angular momentum parameter ${\bm l}$ of the two original binary components.\footnote{ $(G/c)m_1m_2\bm{l}$ is the orbital angular momentum at the ISCO} 
We adopt the formula obtained from numerical simulations of BBH mergers in \citet{Rezzolla08},
\begin{eqnarray}
\label{eq:af_merger}
{\bm a}^\mathrm{f}=\frac{1}{(1+q)^2}
({\bm a}_1+{\bm a}_2 q^2+{\bm l} q),
\end{eqnarray}
where $q\equiv m_2/m_1\leq 1$ is the mass ratio.
The magnitude of ${\bm l}$ is given by 
\begin{align}
|{\bm l}|= &\frac{s_4}{(1+q^2)^2}(|{\bm a}_1|^2+|{\bm a}_2|^2q^4 + 2 |{\bm a}_1||{\bm a}_2|q^2 \cos \theta_{12})
\nonumber \\
&+\left( \frac{s_5 \eta + t_0 + 2}{1+q^2}\right) (|{\bm a}_1|\cos \theta_{1b} + |{\bm a}_2|q^2\cos \theta_{2b})\nonumber \\
&+2\sqrt{3}+ t_2 \eta + t_3 \eta^2
\end{align}
where $\eta\equiv q/(q+1)^2$ is the symmetric mass ratio,  $s_4=-0.129$, $s_5=-0.384$, $t_0=-2.686$, $t_2=-3.454$, and $t_3=2.353$ are values obtained in \citet{Rezzolla08}, and
$\theta_{12}$, $\theta_{1b}$, and $\theta_{2b}$ are the angles between the spins of the two BHs and their orbital angular momentum, 
\begin{equation}
    \cos \theta_{12} = \frac{{\bm a}_1 \cdot {\bm a}_2}{|{\bm a}_1||{\bm a}_2|}, 
\end{equation}
\begin{equation}
    \cos \theta_{1b} = \frac{{\bm a}_1 \cdot {\bm l}}{|{\bm a}_1||{\bm l}|}, 
\end{equation}
\begin{equation}
    \cos \theta_{2b} = \frac{{\bm a}_2 \cdot {\bm l}}{|{\bm a}_2||{\bm l}|}.
\end{equation}
According to Eq.~\eqref{eq:af_merger}, if $a_1=a_2=0$, $a^\mathrm{f}$ monotonically increases from 0.58 to 0.69 as $q$ increases from $0.4$ to $1$ and decreases to 0 as $q\rightarrow0$. 

Following \citet{Rezzolla08} and \citet{Dubois14}, we assume that GW radiation does not affect the direction of the orbital angular momentum, 
and set the direction of ${\bm l}$ to the binary orbital angular momentum 
at merger. 
Although \citet{Barausse09} suggested that GW radiation modifies the binary orbital angular momentum direction just before merger, we neglect this correction for simplicity as results are not sensitive to the precise ${\bm a}^f$ direction. 
Similarly, we assume that $\chi_\mathrm{eff}$ values are unaffected by relativitistic effects. This is justified if the BH spin directions and the orbital angular momentum directions are not influenced by the effects before binaries enter the frequency above which the LIGO/Virgo detectors are sensitive, which is $\sim 10\,\mathrm{Hz}$.  
The orbital angular momentum direction is not directly influenced by GW radiation reaction significantly for circular orbits at lower frequencies \citep{Barausse09}.
The BH spin directions can be systematically affected by spin-orbit resonances due to precession of the BH spins, which can align or anti-align the BH spins with each other or cause nutation \citep{Kesden10,Gerosa19}. In our simulations, such resonances or nutation occur only within the detectable frequency band above $10\,\mathrm{Hz}$ for $\gtrsim95\%$ of stellar BH mergers. Thus, we conclude that 
these general relativistic effects have a small impact on the detectable $\chi_\mathrm{eff}$ distribution for LIGO/VIRGO.

\subsection{Evolution of binary orbital angular momentum direction}
\label{sec:am_ev}

\subsubsection{Initial orbital angular momentum direction}

We consider four types of BH binaries 
distinguished by their 
formation process: (i) pre-existing binaries, (ii) gas-capture binaries, (iii) dynamically formed binaries, and 
(iv) remnants of stellar binaries formed in-situ. 
For binaries belonging to (iii) 
we set $\hat{\bm J}_\mathrm{bin}=\pm \hat{J}_\mathrm{AGN}$, 
and the ratio of aligned binaries over anti-aligned binaries to 1, as suggested by simulations of binary formation in migration traps (Secunda et al. in prep). 
For binaries belonging to (iv), we 
also set the ratio to be 1 for simplicity, although we note that this ratio is highly uncertain. 
On the other hand, for binaries belonging to (i) or (ii),
the orbital angular momentum directions are presumed to be random. For simplicity, we assume that all binaries have zero eccentricity. 
We expect that this assumption does not significantly change the evolution of binary orbital angular momenta or BH spins.

As we show below, the initial angular momentum directions of binaries 
have a relatively small effect on
the $\chi_\mathrm{eff}$ distribution measured at merger, because these directions are frequently randomized by binary-single interactions.

\subsubsection{Gas accretion}
\label{sec:gas_acc_am}

The orbital angular momentum directions of binaries evolve due to accretion torques. 
If the circumbinary gas is rotating in the same direction as the AGN disk, the
binaries are aligned with the AGN disk since the relative velocity between 
the binary components and the gas is reduced by the accretion. 
Referring to \citet{Lubow99}, we assume that the angular momentum direction of the captured gas with respect to the binary ($\hat {\bm J}_\mathrm{gas}$) is the same as the angular momentum direction of the AGN disk with respect to the SMBH ($\hat{\bm J}_\mathrm{AGN}$) 
for binaries embedded in the AGN disk. 
The angular momentum of the captured gas is added to the binary orbital angular momentum as ${\bm J}_\mathrm{bin}^\mathrm{f}={\bm J}_\mathrm{bin}+{\bm J}_\mathrm{gas}$, where we set 
\begin{equation}
{\bm J}_\mathrm{gas}=f_\mathrm{rot} s v_{\mathrm{bin}}(s){\dot m}_{\mathrm{BHL}}\Delta t \hat{\bm J}_\mathrm{AGN}, 
\end{equation} 
and where $v_{\mathrm{bin}}(s)=\sqrt{G m_{\mathrm{bin}}/s}$ is the relative rotation velocity of binary components, and $f_\mathrm{rot}$ is a parameter determining the efficiency of the alignment of the binary angular momentum direction due to gas capture. 
In the fiducial model, we set $f_\mathrm{rot}=1$ assuming that the binary receives a torque from gas circularly rotating at $\sim s$ from the binary.
However, a low degree of rotation ($f_\mathrm{rot}\sim 0$) of gas accreting onto a low-mass object in an AGN disk is suggested in simulations by \citet{Baruteau11} and \citet{Derdzinski18}. For completeness, we investigate this case, as well as an opposite extreme case with $f_\mathrm{rot}=10$ (models~M13~and~M14). 
Note that during gas accretion, the binary separation also evolves due to type I/II torque of the circumbinary disk (Paper~I).

\subsubsection{Binary-single interaction} 

\label{sec:bs_am}

After a hard binary-single interaction, the orbital angular momenta of binaries are modified due to chaotic interactions. 
In this paper, we simply assume that after a hard binary-single interaction, the orbital angular momentum direction of a binary becomes isotropically random. 

Whenever a binary-single interaction occurs in the simulation, we choose a nearby third object, and assign a recoil kick velocity to it. 
If this third object is itself a binary, we assume that the softer binary is disrupted, while the harder binary experiences a hard binary-single interaction by regarding the softer binary as a single object for simplicity, and assign the recoil kick velocity to its center of mass.

\subsection{Merger prescription}

Since we track the evolution of the BH spins and the binary orbital angular momenta, we can estimate the recoil velocity due to  anisotropic GW radiation and mass loss at mergers more precisely. We add the following prescriptions to the model used in Paper~I. 

\subsubsection{Recoil velocity at merger}

Due to the burst of anisotropic GW radiation at merger, a remnant BH receives a recoil kick. 
To calculate the recoil velocities,
we adopt the fitting formulae obtained from numerical simulations by \citet{Lousto12},
\begin{align}
    {\bm v}_\mathrm{GW} =& v_\mathrm{m} \hat{\bm e}_x + v_\perp (\cos \xi \hat{\bm e}_x + \mathrm{sin} \xi \hat{\bm e}_y) + v_{\parallel} \hat{\bm e}_z,\nonumber\\
    v_\mathrm{m} =& A \eta^2 \sqrt{1-4\eta}(1+B\eta), \nonumber\\
    v_\perp =& H \frac{\eta^2}{(1+q)}(a_2^\parallel - q a_1^\parallel), \nonumber\\
    v_\parallel =&\frac{16\eta^2}{1+q}
    \left[V_{1,1}+V_A \tilde{S}_{\parallel} +V_B {\tilde{S}_{\parallel}}^2 +V_C {\tilde{S}_{\parallel}}^3 \right]
    \nonumber\\
    &\times|a_1^\perp - q a_2^{\perp} |\cos (\phi_{\Delta} - \phi_1),
\end{align}
where $v_\mathrm{m}$ is a mass-asymmetry contribution, $v_\perp$ and $v_\parallel$ are kick components perpendicular and parallel to the orbital angular momentum, respectively,
\begin{equation}
\tilde{\bm S}\equiv2\frac{{\bm a}_1+q^2 {\bm a}_2}{(1+q)^2},
\end{equation}
$\hat{\bm e}_x$, $\hat{\bm e}_y$ are orthogonal unit vectors in the orbital plane, and $\hat{\bm e}_z$ is the direction of the binary orbital angular momentum.  
The symbols $\parallel$ and $\perp$ refer to the directions parallel and perpendicular to the orbital angular momentum, respectively, and the numerical constants are 
$A=1.2 \times 10^4\,\mathrm{km/s}$, $B=-0.93$, $H=6.9  \times 10^3\,\mathrm{km/s}$, $\xi = 145^\circ \pm 5^\circ$, 
$V_{1,1}=3678\,\mathrm{km/s}$, 
$V_A=2481\,\mathrm{km/s}$, 
$V_B=1792\,\mathrm{km/s}$, 
and $V_C=1507\,\mathrm{km/s}$, 
$\phi_1$ is the phase angle of the binary, and 
$\phi_{\Delta}$ is the angle between the in-plane component of the vector 
\begin{equation}
    {\bm \Delta}=m_\mathrm{bin}^2 \frac{{\bm a}_1 - q{\bm a}_2}{1+q} 
\end{equation}
and the infall direction at merger. 
We choose $\phi_{\Delta}-\phi_1$ from a random distribution uniform 
in $[0,\pi]$.

\subsubsection{Mass loss at merger}

Due to GW radiation, some fraction of the BH mass is radiated away. 
We adopt a simplified approximation for the remnant mass~from \citet{Barausse12},
\begin{align}
\label{eq:mrem}
    \frac{m_\mathrm{rem}}{m_\mathrm{bin}}=
    1-\eta (1-4\eta)
    [1-E_\mathrm{ISCO}(\tilde{a}_\mathrm{\parallel})]\nonumber\\
    -16\eta^2
    [p_0+4p_1 \tilde{a}_\mathrm{\parallel}(\tilde{a}_\mathrm{\parallel}+1)]],
\end{align}
where 
\begin{equation}
\tilde{E}_\mathrm{ISCO}(\tilde{a}_\mathrm{\parallel})\equiv\left(1-\frac{2}{3r_\mathrm{ISCO}(\tilde{a}_\mathrm{\parallel})}\right)^{1/2}
\end{equation}
is the energy per unit mass of a particle with spin 
\begin{equation}
\label{eq:spin_eff}
\tilde{\bm a}=\frac{{\bm a}_1+q^2 {\bm a}_2}{(1+q)^2},
\end{equation}
and $p_0=0.04827$ and $p_1=0.01707$ are parameters obtained by fitting numerical results.

\subsubsection{Merger condition}

We assume that a binary merges when its
separation $s$ 
becomes smaller than the ISCO of 
a particle with mass $m_{\rm bin}$ and spin given in Eq.~\eqref{eq:spin_eff}. 
We subsequently treat the object as a single BH with a mass given by Eq.~(\ref{eq:mrem}).

\begin{table*}
\begin{center}
\caption{Fiducial values of our model parameters.}
\label{table_parameter_model}
\hspace{-5mm}
\begin{tabular}{c|c}
\hline 
Parameter & Fiducial value \\
\hline\hline
Initial BH spin magnitude & $|{\bm a}|=0$\\\hline
Angular momentum directions of circum-BH disks & $\hat{{\bm J}}_\mathrm{CBHD}=\hat{{\bm J}}_\mathrm{AGN}$ for single BHs,\\&
$\hat{{\bm J}}_\mathrm{CBHD}=\hat{{\bm J}}_\mathrm{bin}$ for BHs in binaries\\\hline
Ratio of viscous parameters & $\nu_2/\nu_1=10$ \\\hline
Efficiency of alignment of ${\bm J}_\mathrm{bin}$ due to gas capture& $f_\mathrm{rot}=1$\\\hline 
Mass of the central SMBH & $M_\mathrm{SMBH}=4\times 10^6\,\Msun$ \\\hline
Gas accretion rate at the outer radius & ${\dot M}_\mathrm{out}=0.1\,{\dot M}_\mathrm{Edd}$\\\hline
Fraction of pre-existing binaries & $f_\mathrm{pre}=0.15$ \\\hline
Power-law exponent for the initial density profile for BHs & $\gamma_{\rho}=0$ \\\hline
Parameter setting the initial velocity anisotropy for BHs & $\beta_\mathrm{v}=0.2$ \\\hline
Efficiency of angular momentum transport in the $\alpha$-disk & $\alpha_\mathrm{SS}=0.1$ \\\hline
Stellar mass within 3 pc &$M_\mathrm{star,3pc}=10^7\,\Msun$\\\hline 
Stellar initial mass function slope & $\delta_\mathrm{IMF}=2.35$\\\hline
Angular momentum transfer parameter in the outer disk &$m_\mathrm{AM}=0.15$\\\hline
Accretion rate in Eddington units onto stellar-mass BHs &$\Gamma_\mathrm{Edd,cir}=1$\\\hline
Numerical time-step parameter &$\eta_t=0.1$\\\hline
Number of radial cells storing physical quantities &$N_\mathrm{cell}=120$\\\hline
Maximum and minimum $r$ for the initial BH distribution&  $r_\mathrm{in,BH}=10^{-4}$ pc, $r_\mathrm{out,BH}=3$ pc \\\hline
\end{tabular}
\end{center}
\end{table*}

\subsection{Numerical choices}
\label{sec:numerical_choice}

Table~\ref{table_parameter_model} lists the parameter values adopted in the fiducial model (the same as Model~1 in Paper~I). 
We assume that stars are distributed spherically with a Maxwell-Boltzmann velocity distribution, 
the stellar mass within 3 pc is $M_\mathrm{star,3pc}=10^7\,\Msun$, 
and the power-law slope of the stellar initial mass function (IMF) is $\delta_\mathrm{IMF}=2.35$. 
BHs are initially distributed from $r_\mathrm{in,BH}=2\times 10^{-4}$ pc to $r_\mathrm{out,BH}=3$ pc 
with a cumulative radial profile
\begin{align}
\label{eq:bh_density}
\frac{dN_{\rm BH, ini}(r)}{dr}
\propto r ^{\gamma_{\rho}}
\end{align}
with a power-law index $\gamma_{\rho}=0$, where $N_{\rm BH, ini}(r)$ labels the total initial number of BHs within a distance $r$ from the central SMBH.  Note that BHs are assumed to have a flattened axisymmetric distribution (see Paper I). 
The $x$, $y$, and $z$ velocities for BHs 
relative to the local Keplerian value $v_\mathrm{Kep}(r)$ 
are initially drawn from a Gaussian distribution with the dispersion 
of $\beta_\mathrm{v} v_\mathrm{Kep}(r)/\sqrt{3}$, where  
$\beta_\mathrm{v}$ is
a velocity anisotropy parameter
set to $\beta_\mathrm{v}=0.2$ 
motivated by vector resonant relaxation \citep{Szolgyen18}. 
The total number and mass in BHs are calculated from the stellar mass, the stellar IMF, and the relation between the stellar and BH mass derived in \citet{Belczynski10}.  
We set the fraction of pre-existing binaries 
to be $f_\mathrm{pre}=0.15$. 
In the fiducial model, there are initially $2\times 10^4$ BHs and $1.5\times 10^3$ binaries.
As in Paper~I, the time step parameter is $\eta_t=0.1$, and 
the number of radial cells storing physical quantities is $N_\mathrm{cell}=120$. 

The mass of the central SMBH is $M_\mathrm{SMBH}=4\times 10^6\,\Msun$, the gas accretion rate from the outer radius 
is ${\dot M}_\mathrm{out}=0.1 {\dot M}_\mathrm{Edd}$, where ${\dot M}_\mathrm{Edd}=L_\mathrm{Edd}/(\eta_\mathrm{c} c^2)$ is the Eddington accretion rate, here defined including a radiative efficiency of $\eta_\mathrm{c}=0.1$. 
The efficiency of angular momentum transport in the $\alpha$-disk is $\alpha_\mathrm{SS}=0.1$ \citep{Shakura73}, 
and the angular momentum transfer parameter is $m_\mathrm{AM}=0.15$ \citep{Thompson05}.

\begin{figure*}
\begin{center}
\includegraphics[width=140mm]{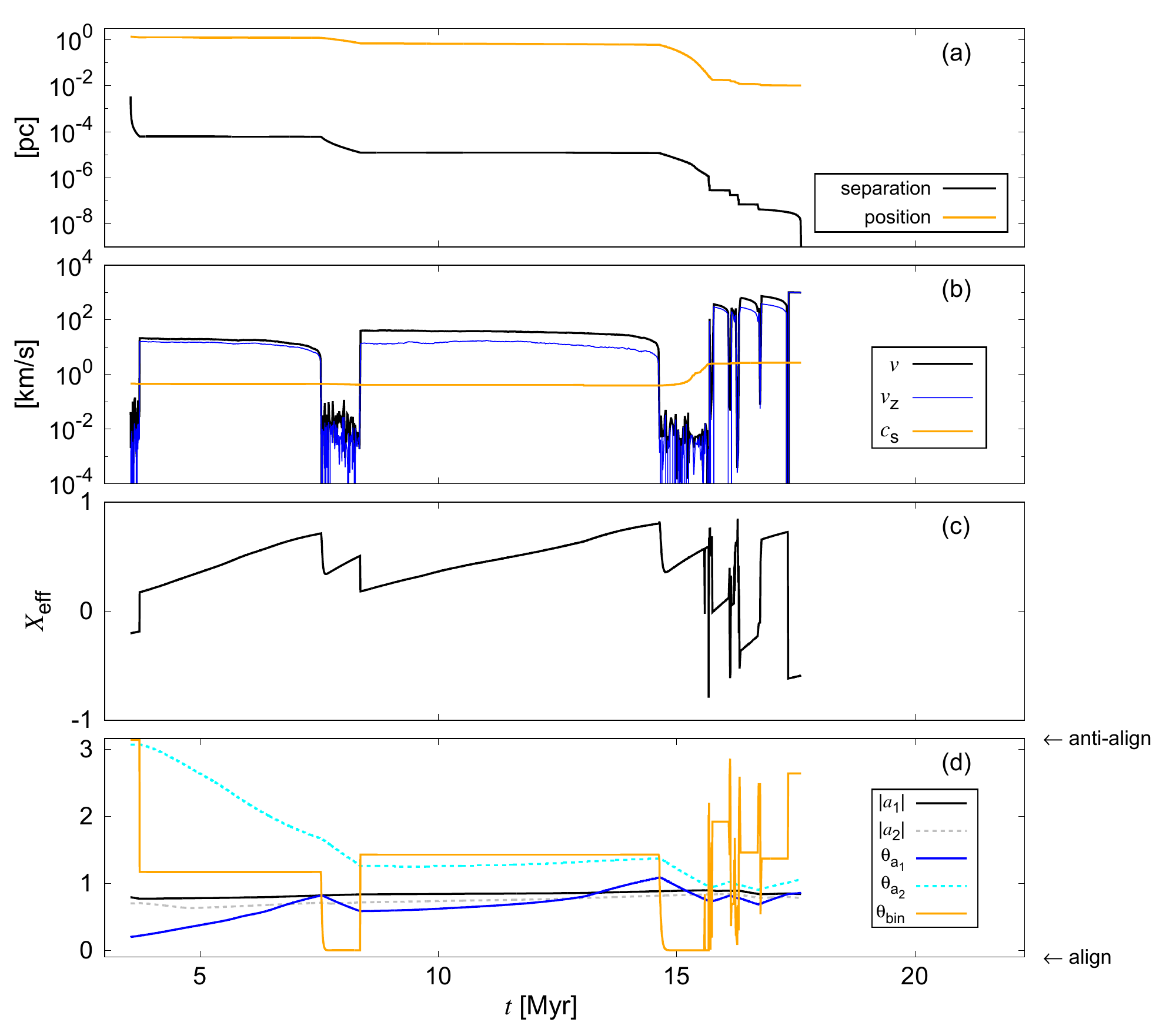}
\caption{
Evolution of several quantities for a binary in model~M6 (in which $a_0=0.7$) formed via the gas-capture mechanism. 
(a): Binary separation (black) and distance from the SMBH (orange). 
(b): Velocity of the center of mass of the binary relative to the local motion of the AGN disk $v$ (black), $z$-direction velocity $v_{\mathrm{z}}$ (blue), and sonic velocity of gas near the binary $c_\mathrm{s}$ (orange). 
While $v_{\mathrm{z}} \leq c_\mathrm{s}$, the typical height of orbital motion for the binary is thinner than the scale height of the AGN disk, which means that
the binary is embedded in the AGN disk. 
(c): The effective spin parameter $\chi_\mathrm{eff}$. 
(d): 
Spin magnitude of binary components ($|a_1|$ and $|a_2|$, black and dashed gray), the angle between the BH spins $\hat{\bm a}_1$ or $\hat{\bm a}_2$ and the AGN angular momentum direction $\hat{\bm J}_\mathrm{AGN}$ (blue and dashed cyan), and the angle between the binary angular momentum direction $\hat{\bm J}_\mathrm{bin}$ and $\hat{\bm J}_\mathrm{AGN}$ (orange). 
}
\label{fig:tot_cont}
\end{center}
\end{figure*}

\section{Results}

\subsection{$\chi_\mathrm{eff}$ evolution: an illustrative example}

We used the combination of semi-analytical calculations and simulations, described above, to investigate the $\chi_\mathrm{eff}$ distribution of BHs merging in AGN disks. An illustrative example of the evolution of such a BH binary is shown in Figure \ref{fig:tot_cont}. 

The binary in this figure forms via the gas-capture mechanism at $t=3.5$~Myr at a distance of 1.3~pc from the SMBH, with an initial separation of $3.4\times 10^{-3}$ pc. The masses of the binary components are $10.7$ and $6.8$ $\Msun$, 
the magnitude of BH spins are 0.79 and 0.70, and the angles between the BH spins and $\hat{\bm J}_\mathrm{AGN}$ are 0.20 and 3.1, respectively.\footnote{At birth the initial value of both BH spins is 0.7 and the BH masses are $10.0$ and $6.8\,\Msun$, but one of the BHs mass and spin evolved due to gas accretion prior to binary formation.}
The separation of the binary (black line in panel~(a)) decreases, successively, due to gas dynamical friction, binary-single interactions, and GW radiation as shown in Paper~I. 
Although a binary is disrupted when the binary separation exceeds the Hill radius, ionization is extremely rare ($\approx1\%$ of the number of mergers) due to rapid hardening by gas dynamical friction in early phases (see Figs.~5 and 6 of Paper~I). 

While the binary is in the AGN disk ($v_{z}<c_\mathrm{s}$, blue and orange lines in panel~(b)), $\hat{\bm J}_\mathrm{bin}$ aligns with $\hat{\bm J}_\mathrm{AGN}$ due to accretion torque (orange line in panel~(d)). Also, $\hat{\bm a}_1$ and $\hat{\bm a}_2$ (blue and cyan lines in panel~(d)) evolve towards the angular momentum direction of a circum-BH disk, which is set to be the same as $\hat{\bm J}_\mathrm{bin}$. 
Such alignment of $\hat{\bm a}_1$ and $\hat{\bm a}_2$ with $\hat{\bm J}_\mathrm{bin}$ increases $\chi_\mathrm{eff}$ (panel~(c)). 
The spin magnitudes, $|a_1|$ and $|a_2|$ evolve due to gas accretion (black and gray lines in panel~(d)), but only by $20\%$ and $11\%$, which are much smaller than the change in $\chi_\mathrm{eff}$. 
Until $t \sim 5$ Myr, since the anti-alignment condition (Eq.~\ref{eq:alignment_king}) is satisfied for the secondary BH, $|a_2|$ slightly decreases as gas accretes.

After each binary-single interaction, $\hat{\bm J}_\mathrm{bin}$ is randomized, which reduces $|\chi_\mathrm{eff}|$ on average. 
Binary-single interactions become very frequent at $\lesssim0.01$ pc due to the high BH density, and binaries merging in the inner regions typically have experienced a larger number ($\approx8$) of binary-single interactions. 
Note that kicked BHs in the inner regions are typically easily re-captured, e.g. within $\sim 0.1$ Myr at $r\sim 0.01$ pc for the kick velocity of $\sim 300\,\mathrm{km/s}$ (see Eq.~(26) and Fig.~10 in Paper~I). 
This binary merges outside the AGN disk $17.6$ Myr after it formed, and its components accrete $2.0$ and $1.7\,\Msun$ until their merger. Since $\hat{\bm a}_1$ and $\hat{\bm a}_2$ are roughly anti-aligned with $\hat{\bm J}_\mathrm{bin}$ following binary-single interactions, $\chi_\mathrm{eff}$ at merger has a negative value of -0.59. 
Thus, $\chi_\mathrm{eff}$ in this case evolves through both gas accretion and binary-single interactions.

\begin{figure*}\begin{center}
\includegraphics[width=180mm]{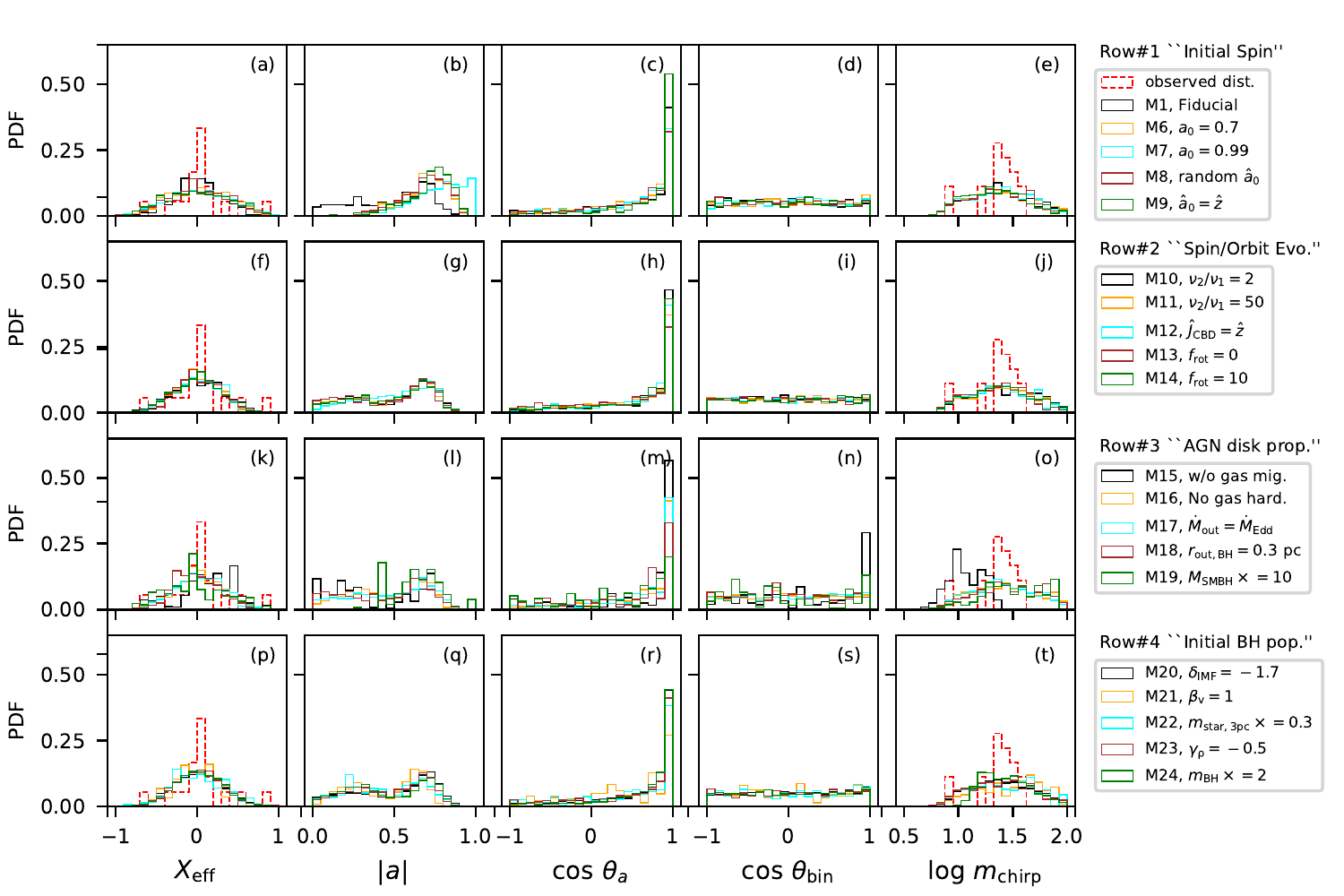}
\caption{
The probability distributions of $\chi_\mathrm{eff}$, $a$, $\cos \theta_a$, $\cos \theta_\mathrm{bin}$, and $m_\mathrm{chirp}$ 
at the time of the merger (left to right columns). 
Dashed red lines in the left-most and right-most panels are the distributions inferred from the GW events observed by LIGO/Virgo (Table~\ref{table_parameter_data}). 
All predicted distributions are weighted by the volume detectable by LIGO, and observational errors  have been added to the predicted $\chi_\mathrm{eff}$ and $m_\mathrm{chirp}$ values (see text). 
The black lines in the first row correspond to the fiducial model (M1), and the other colors and the other three rows consider different types of model-variations (M2--M24), as labeled on the right, and listed in Table~\ref{table_results}.
}\label{fig:spin_dist}
\end{center}\end{figure*}

\begin{figure*}\begin{center}
\includegraphics[width=180mm]{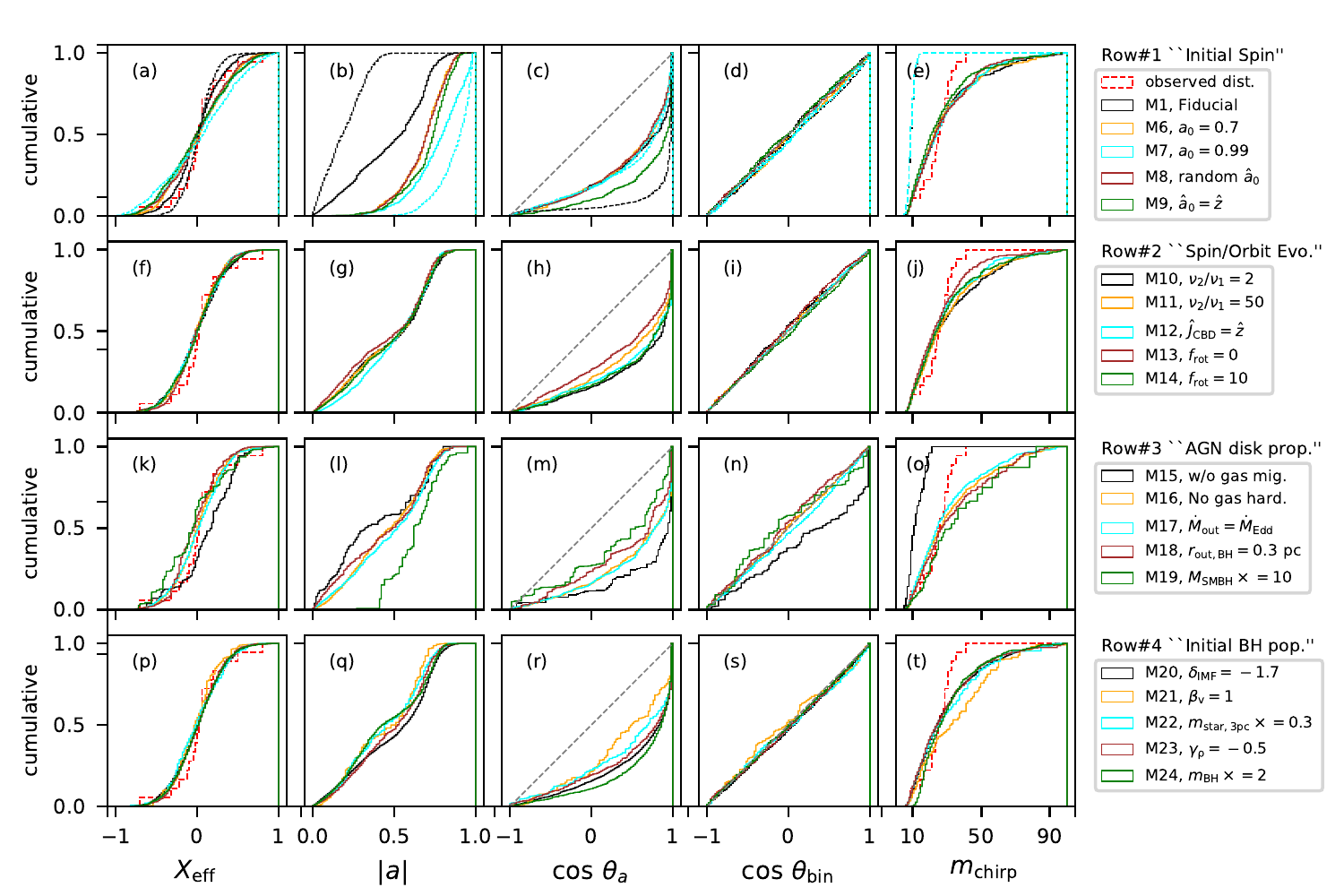}
\caption{
Same as Fig.~\ref{fig:spin_dist}, but shows cumulative distributions. 
The dashed gray lines in the third and fourth columns represent isotropic distributions. 
In models~M1 and M7, we present the distributions of first generation mergers by dashed lines in panels~(a)-(e). 
}\label{fig:spin_hist_dist}
\end{center}\end{figure*}

\begin{figure*}\begin{center}
\includegraphics[width=195mm]{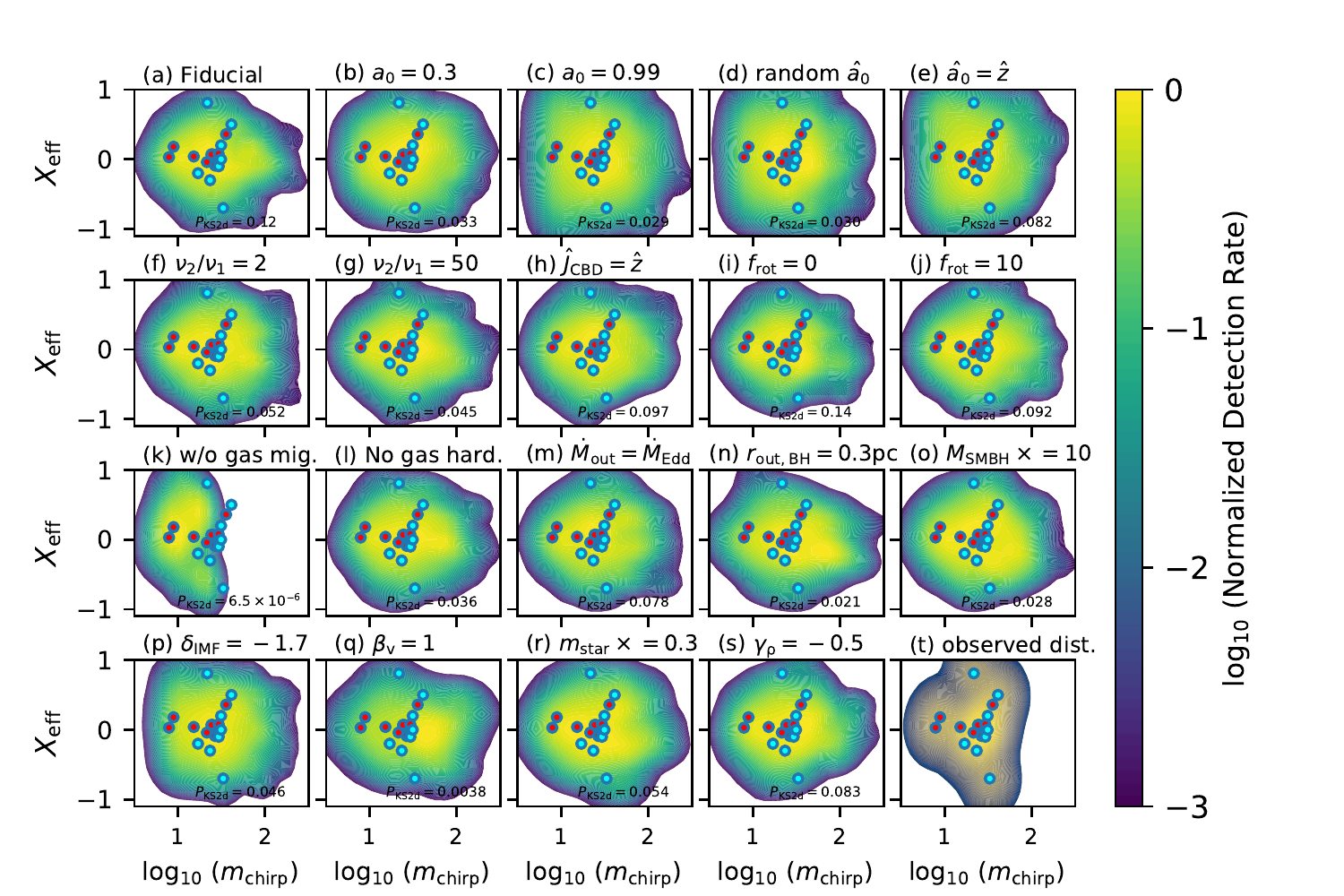}
\caption{
The normalized detection rate of mergers in $m_\mathrm{chirp}$ vs. $\chi_\mathrm{eff}$ plane for several models. 
The detection rate is smoothed by performing a kernel-density estimate, and normalized by the maximum value in each plane. 
Observational errors have been added to the predicted $\chi_\mathrm{eff}$ and $m_\mathrm{chirp}$ values. 
Panels~(a-s) shows the detection rate distribution at 10 Myr for models~M1, M4, M7--M23, respectively. 
The probability that the $\chi_\mathrm{eff}$ and $m_\mathrm{chirp}$ distributions for all events is reproduced by each model (the AGN contribution to all merger is $f_\mathrm{AGN}=1$) estimated by the KS test ($P_{\mathrm{KS},\chi_\mathrm{eff},m_\mathrm{chirp}}$) is shown in a lower right corner.
Panel~(t) shows the distribution derived by performing a kernel-density estimate for the observed distribution, which is presented by different colors to emphasize its peculiarity. 
The values for the events inferred from the LIGO/Virgo collaboration (red circles) and the IAS group (cyan circles) are overplotted in all panels. 
}\label{fig:mc_xeff_dist}
\end{center}\end{figure*}

\begin{figure}\begin{center}
\includegraphics[width=95mm]{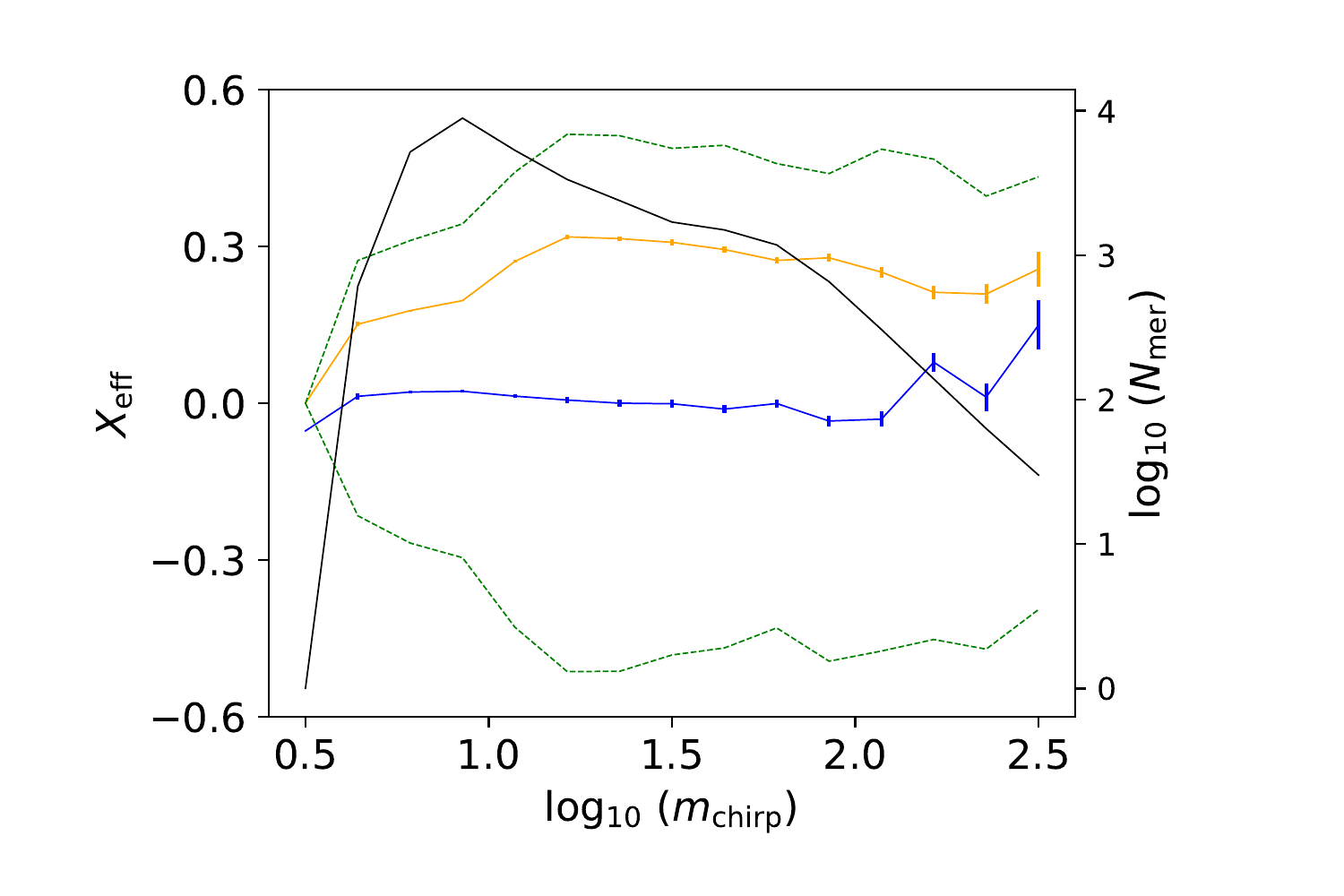}
\caption{
The average values and one sigma errors of the mean (cyan) and the standard deviation (orange) of $\chi_\mathrm{eff}$ as a function of $m_\mathrm{chirp}$ in ten additional simulations of model~M1 with independent realizations of the initial condition. 
Dashed green lines represent the 5th and 95th percentiles of $\chi_\mathrm{eff}$, respectively. 
Black line shows the number of mergers summed over 11 runs in mass bins of 0.133 dex (with corresponding values shown on the y-axis on the right). 
}\label{fig:mc_xeff_fid}
\end{center}\end{figure}

\begin{figure}\begin{center}\includegraphics[width=95mm]{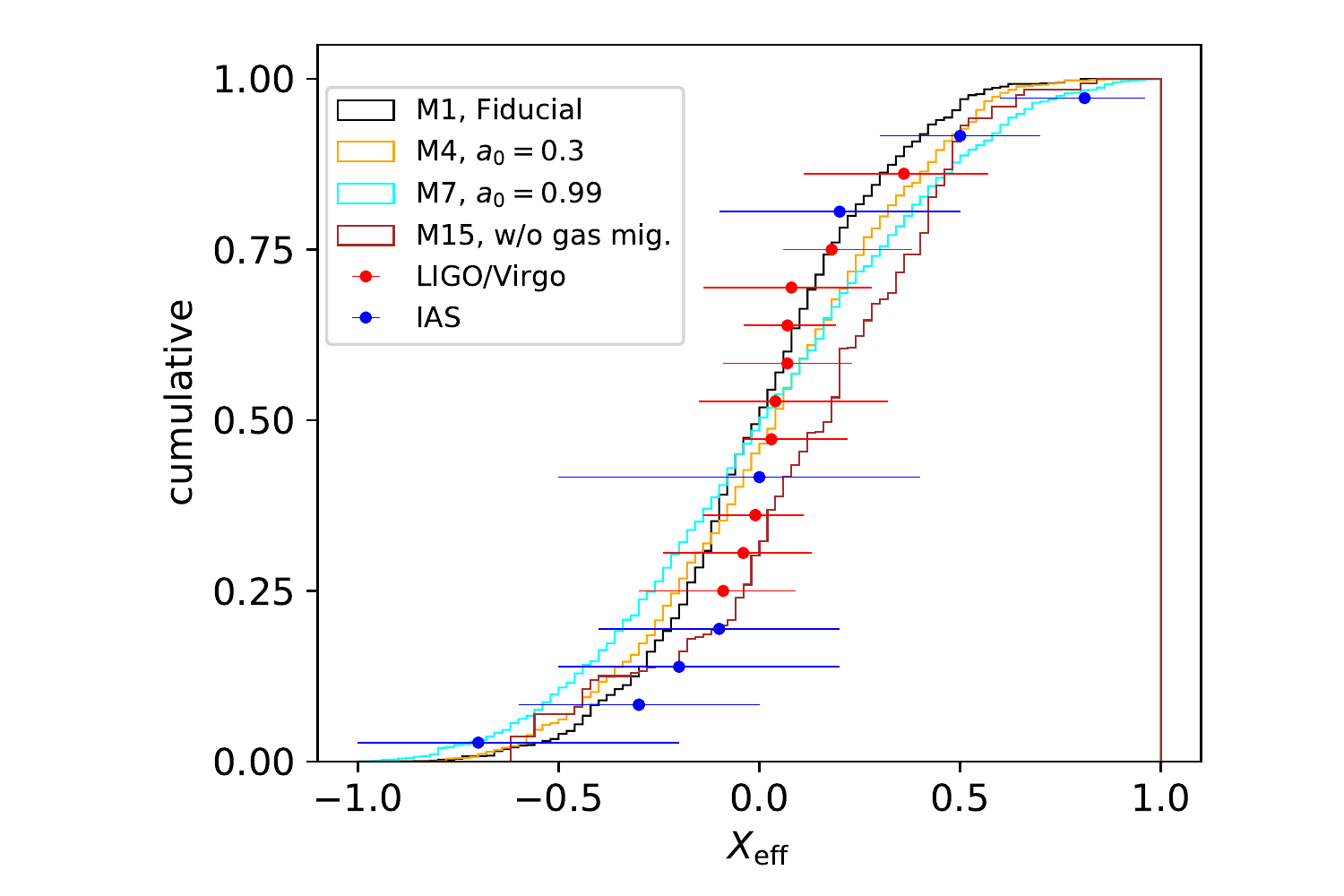}\caption{
Comparison between the $\chi_\mathrm{eff}$ distribution inferred from the observed GW events and those predicted in our models. 
Black, orange, cyan, and brown lines show the distribution in models~M1, M4, M7, and M15, respectively. The predicted distributions are weighted by the detectable volume, 
and include observational errors. 
Red and blue circles show the median $\chi_\mathrm{eff}$ values reported by \citet{TheLIGO18} and the IAS group \citep{Zackay19,Zackay19b,Venumadhav19}, respectively. Error bars correspond to $90\%$ credible intervals. 
}\label{fig:spin_obs}\end{center}\end{figure}

\begin{figure}\begin{center}\includegraphics[width=95mm]{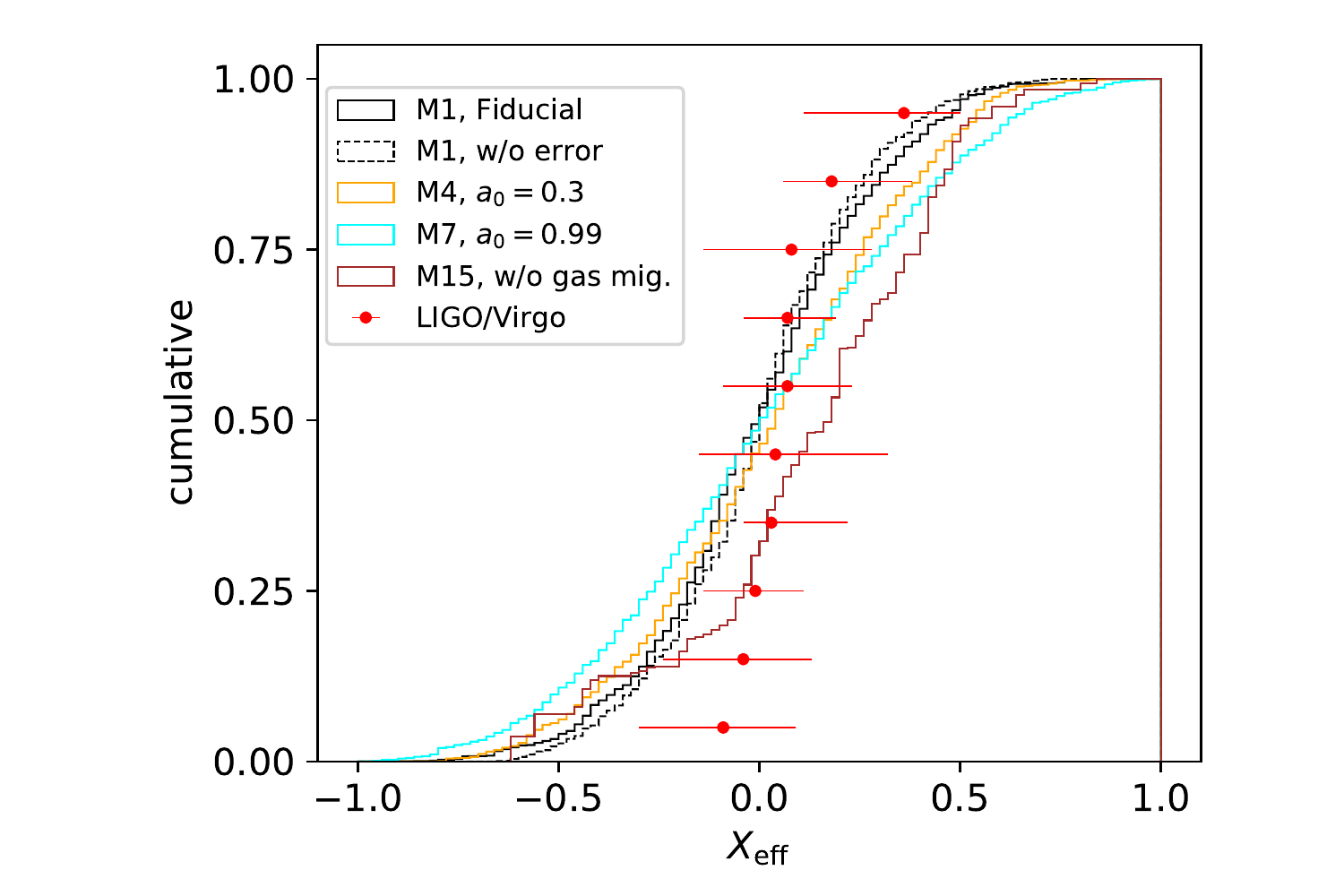}\caption{
Same as Fig.~\ref{fig:spin_obs}, but compared only with the events reported by the LIGO and Virgo collaborations (O1/O2). 
The distribution without errors are shown by dashed line. 
}\label{fig:spin_obs_ligo}\end{center}\end{figure}

\begin{figure}\begin{center}
\includegraphics[width=80mm]{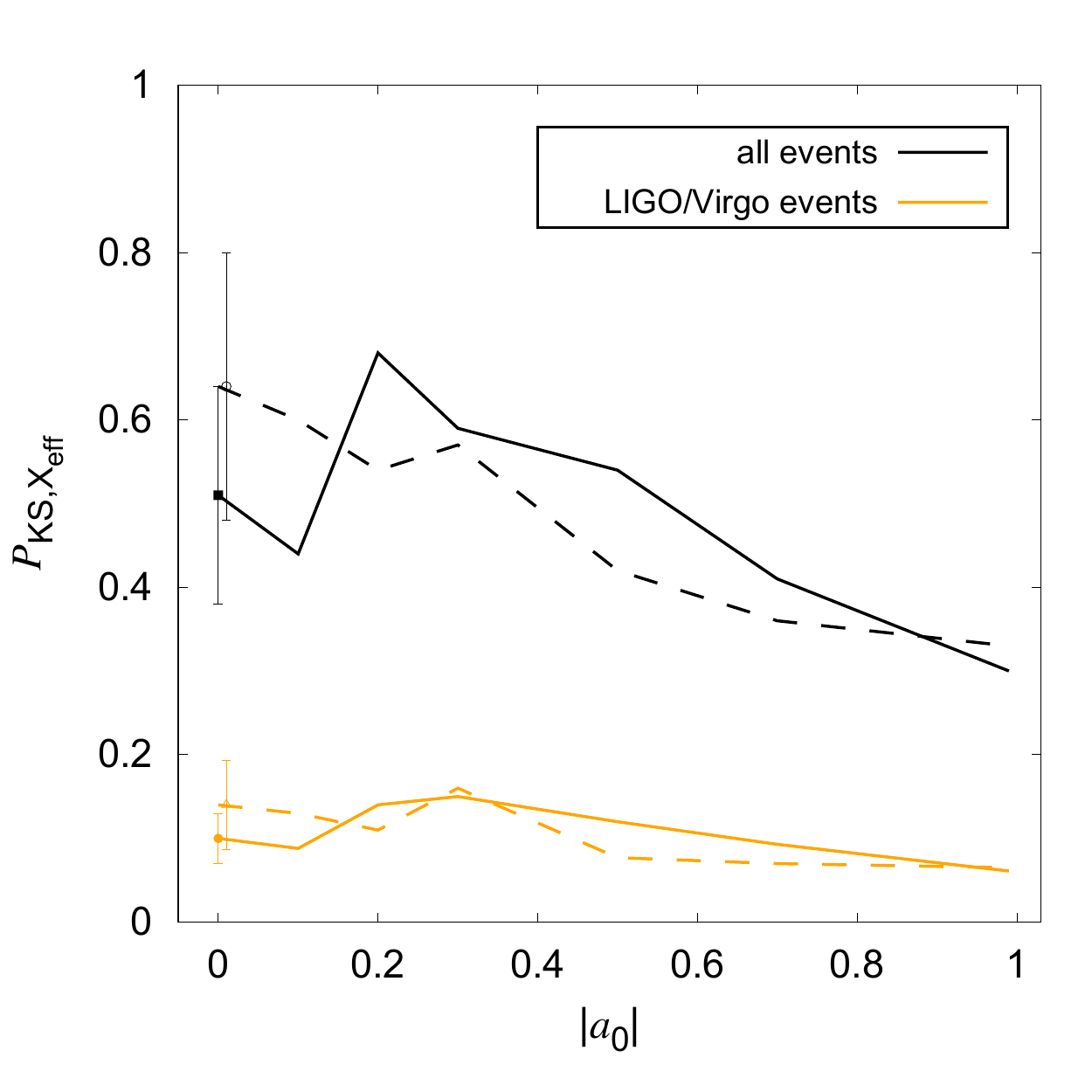}
\caption{The KS probabilities that the one-dimensional $\chi_\mathrm{eff}$ distribution inferred from 
observed GW events 
is consistent with models assuming different initial spin magnitudes $|a_0|$ (i.e. models M1--M7).  
Black and orange lines show the results in which all observed GW events and only the LIGO/Virgo events are used, respectively. Solid and dashed lines show the results in which errors are and are not included in the predicted $\chi_\mathrm{eff}$ distribution, respectively. 
For $a_0=0$, the errors and means are calculated by performing ten additional runs with different realizations of the initial conditions. 
The observed $\chi_\mathrm{eff}$ distribution slightly favors moderate values of $|a_0|\lesssim 0.5$. 
}
\label{fig:px_a0}
\end{center}\end{figure}

\begin{figure*}\begin{center}
\includegraphics[width=120mm]{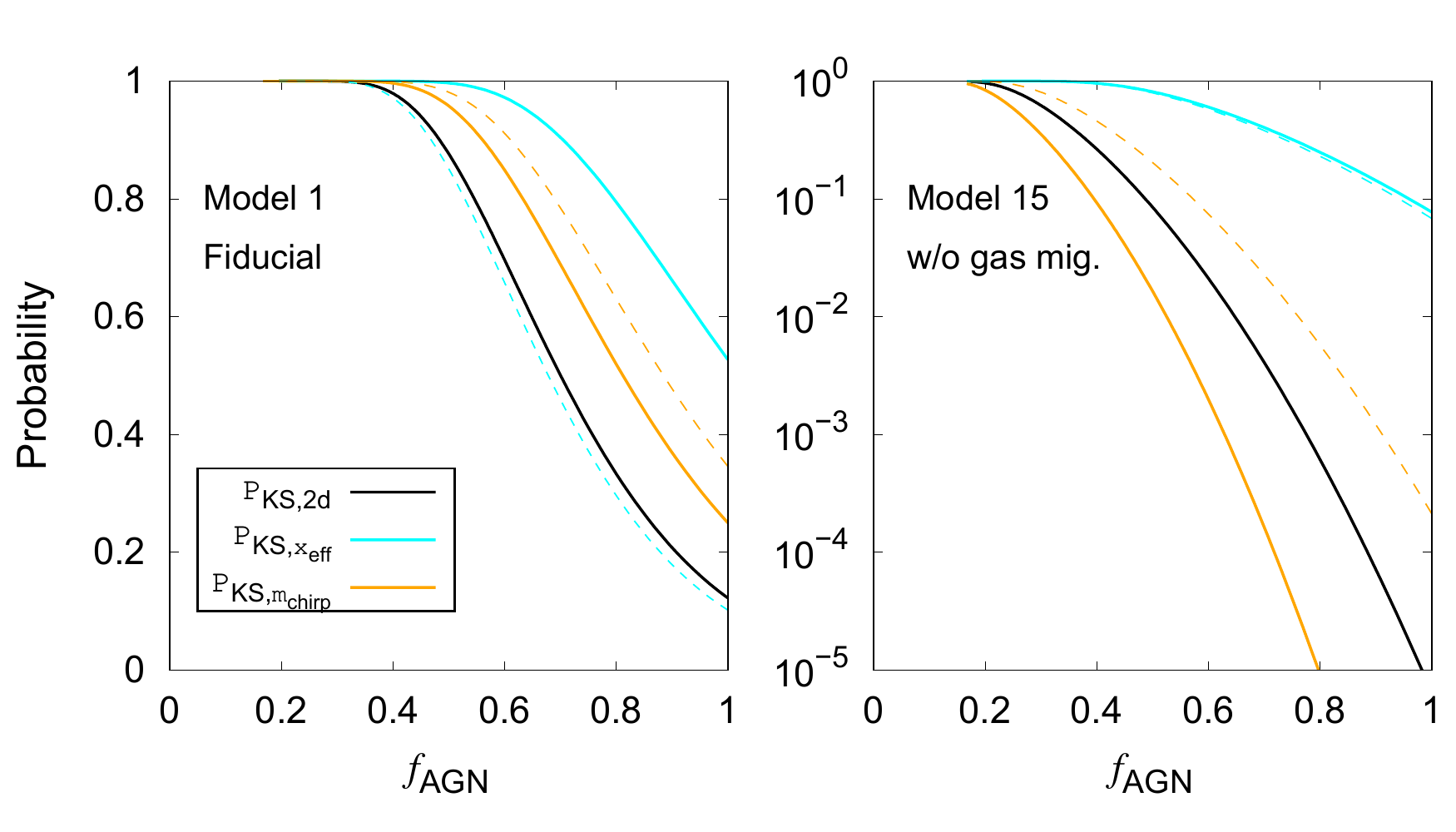}
\caption{
The KS probabilities that all events are produced by each model, as a function of the fraction of all mergers produced in AGN disks. 
Black, cyan, and orange lines show 
$P_{\mathrm{KS},\chi_\mathrm{eff},m_\mathrm{chirp}}$, $P_{\mathrm{KS},\chi_\mathrm{eff}}$, and $P_{\mathrm{KS},m_\mathrm{chirp}}$, respectively. 
The left and right panels present results for the fiducial model M1, and the worst-fit model M15, respectively. 
Solid lines show the KS probabilities including all events claimed to date (Table~\ref{table_parameter_data}), while dashed lines include only the LIGO/Virgo O1/O2 events \citep{TheLIGO18}. 
The probabilities remain high ($\gtrsim70\%)$ in all cases, as long as the AGN channel is responsible for $\sim30\%$ or less of all events.
}\label{fig:cont_ks}
\end{center}\end{figure*}

\subsection{$\chi_\mathrm{eff}$ distribution in different models}
\label{sec:dependence}

In this section, we present the probability distribution of $\chi_\mathrm{eff}$ 
as well as of several different quantities at the time of merger, obtained in our models. 
Because of the flexibility of our model, we are also able to study the dependence of these results on the choice of prescriptions and parameter values.   
In Table~\ref{table_results}, we list
the model variations we have investigated.
These include the fiducial model (M1), and 23 different variations (models~M2--M24). The variations can be divided into four categories.  First, 
we examine different choices of the initial BH spin magnitudes and directions (models~M2--M9).
Second, we vary the prescriptions for the evolution of the BH spin directions and the binary angular momentum directions during gas accretion  (models~M10--M14).  Third, we study different parameters related to the AGN disk (models~M15--M19).  Finally, we vary the properties of the initial BH population (models~M20--M24).

Fig.~\ref{fig:spin_dist} shows our main results, namely 
the differential probability distributions of $\chi_\mathrm{eff}$, $|{\bm a}|$, $\cos \theta_a$, $\cos \theta_\mathrm{bin}$, and $m_\mathrm{chirp}$ (first to fifth column, respectively).  
Here, $|{\bm a}|$ stands for either $|{\bm a_1}|$ or $|{\bm a_2}|$ 
and $\theta_a$ and $\theta_\mathrm{bin}$ are the angles between
($\hat{\bm a}$) and the binary orbit ($\hat{\bm J}_\mathrm{bin}$) with respect to the AGN disk ($\hat{\bm J}_\mathrm{AGN}$).
The four different rows in this figure
correspond to the four different types of model variations, as discussed above, and labeled in the figure.

The distributions of all predicted quantities  in Fig.~\ref{fig:spin_dist} are weighted by the detectable volume, which enables us to compare our predictions to the observed distribution (dashed red lines). The detectable volume is calculated via Eq.~(6) of \citet{TheLIGO12} and using the noise spectral density of the ER13 (prior to O3) run of LIGO Hanford \citep{Kissel18}, in which the volume is assumed to depend only on the masses of the binary components. The volume is roughly proportional to $m_1^{2.2}$ for $m_\mathrm{bin}\lesssim 100\,\Msun$ \citep[e.g.][]{Fishbach17}. 
Furthermore, to compare with the observed distributions of $\chi_\mathrm{eff}$ and $m_\mathrm{chirp}$, we add observational errors to $\chi_\mathrm{eff}$ and $m_\mathrm{chirp}$ for each merger in the simulations.\footnote{We note that $|a_1|$, $|a_2|$, $\cos \theta_{a1}$, $\cos \theta_{a2}$ have very large observational uncertainties, these quantities are currently not measurable independently from GW data. We show the predicted distributions without observational errors for these parameters.} For a simple treatment, we draw the errors of $\chi_\mathrm{eff}$ and $m_\mathrm{chirp}$ from independent Gaussian distributions whose $90\%$ intervals are $\pm 0.2$ and $\pm 0.08 m_\mathrm{chirp}$, respectively, which match the typical O1/O2 observational error magnitudes~\citep{TheLIGO18}. These errors are added to our predictions for the analysis of the Kolmogorov-Smirnov (KS) test and Figs.~\ref{fig:spin_dist}--\ref{fig:cont_ks} below. 

In Fig.~\ref{fig:spin_hist_dist} we present the cumulative, rather than the differential probability distributions shown in Fig.~\ref{fig:spin_dist}. 
We present and discuss results at $t=10$ Myr unless stated otherwise (but see Table~\ref{table_results} below for different choices).

The fiducial model is shown by black lines in all of the panels, (a)-(e), of Figs.~\ref{fig:spin_dist} 
and \ref{fig:spin_hist_dist}.   
Panel (d) shows that $\theta_\mathrm{bin}$ represents an isotropic distribution (uniform in $\cos \theta_\mathrm{bin}$). This is because $\hat{\bm J}_\mathrm{bin}$ is frequently randomized by binary-single interactions. 
On the other hand, 
$\hat{\bm a}_1$ and $\hat{\bm a}_2$ 
tend to align with $\hat{\bm J}_\mathrm{AGN}$ (small $\theta_a$) for the following reason. First,
$\hat{\bm a}_1$ and $\hat{\bm a}_2$ are gradually aligned with $\hat{\bm J}_\mathrm{bin}$ due to gas accretion.
However $\hat{\bm J}_\mathrm{bin}$ aligns with $\hat{\bm J}_\mathrm{AGN}$ when the binary is in the AGN disk and it is mostly random when outside of it. Thus, the $\theta_a$ distribution is influenced by the fraction of the time that binaries typically spend in the AGN disk.

The mean spin magnitude $|{\bm a}|$ evolves from 0 in the fiducial model to typically lie in the range $0.08$--$0.64$, $\chi_\mathrm{eff}$ and $m_\mathrm{chirp}$ are typically in the range $-0.22$ to $+0.24$ ($-0.18$ to $+0.21$ if no errors are added) and $8$--$49\,\Msun$, respectively (enclosing 68\% of the total probability). 
By comparison, for first-generation\footnote{mergers among BHs that did not merge earlier} mergers $|{\bm a}|$, $\chi_\mathrm{eff}$, and $m_\mathrm{chirp}$ are distributed in the range $0.044$--$0.32$, $-0.15$ to $+0.20$, and $7$--$11\,\Msun$, respectively. 
The median radial position $r$ for first-generation mergers is 0.012 pc, while that for higher-generation mergers is 0.0098 pc. The weak dependence of the merger location on the generation is because gaps form around BHs at $r\sim 0.01$ pc, which significantly slows down the migration of BHs. 
Also, at $t=0.1$ and 1 Myr, $m_\mathrm{chirp}$ is distributed over the range $6$--$11$ and $7$--$15~\Msun$, respectively, 
and $|{\bm a}|$ is distributed over the range $0.002$--$0.005$ and $0.009$--$0.60$. The spin magnitude $|{\bm a}|$ goes to $\sim0.6$ when the disk is allowed to be present for 1 Myr, because second-generation mergers start between 0.1 and 1 Myr (and for $< 0.1$ Myr, $|{\bm a}|$ remains close to 0). 
Hence, $|{\bm a}|$ and $m_\mathrm{chirp}$ both evolve significantly due to mergers. This trend is also visible in Fig.~\ref{fig:mc_xeff_dist_disp}, which shows that the standard deviation of $\chi_\mathrm{eff}$ increases with $m_\mathrm{chirp}$ (orange line in panel~(a)). It is notable that a similar trend is also seen in the observed distribution (orange line in panel~(t)).

To see the robustness of this trend, in Fig.~\ref{fig:mc_xeff_fid} we plot the weighted mean and standard deviation of $\chi_{\rm eff}$ as a function of mass for ten additional simulations for independent realizations of the initial condition. 
The orange line in Fig.~\ref{fig:mc_xeff_fid} shows the standard deviation averaged over eleven runs. 
Here, the weighted standard deviation of quantity $x_i$ for each model is calculated by
\begin{align}
    s_j=
   \left( \frac{M_j}{M_j-1}
   \sum_{i\in j}w_i (x_i-\bar{x}_{j})^2/\sum_{i\in j}w_i
   \right)^{1/2}
\end{align}
where $i$ is the index of a merger, $w_i$ is the detectable volume, 
$j$ is the index of a bin, 
$M_j$ is the number of nonzero weights, and $\bar{x}_{j}=\sum_{i\in j}w_i x_i/\sum_{i\in j}w_i$ 
is the weighted mean of $x_i$ in the $j^{\rm th}$ bin. 
In Fig.~\ref{fig:mc_xeff_fid}, 
the errors of the mean and the standard deviation are calculated by 
\begin{align}
    \sigma (\bar{x}_{j})=\frac{s_j}{\sqrt{M_j}},
\end{align}
and
\begin{align}
\label{eq:err_stdev}
    \sigma (s_{j})=s_j \left(\frac{1}{2(M_j-1)}\right)^{1/2},
\end{align}
respectively \citep{Harding14}. 
Eq.~\eqref{eq:err_stdev} is approximately correct for $M_j\gtrsim 10$. 
Fig.~\ref{fig:mc_xeff_fid} shows that the standard deviation of $\chi_{\rm eff}$ robustly increases up to $m_\mathrm{chirp}\sim 20\,\Msun$, while it is roughly constant in the range $20 \lesssim m_\mathrm{chirp}\lesssim 100$.

Next, we present the dependence of the distributions on the assumed initial BH spins 
(first row in Fig.~\ref{fig:spin_dist}). 
In models~M1 and M7, the initial BH spins are set to $|a_0|=0$ and 0.99, respectively. 
We also examined five intermediate cases, with $|a_0|=0.1$, $0.2$, $0.3$, $0.5$, and $0.7$  (models~M2--M6) and found that the resulting distributions for all five quantities lie in-between those of the extreme models~M1 and M7. For clarity, the intermediate cases (M2--M5) are therefore not shown in Figs.~\ref{fig:spin_dist}~and~\ref{fig:spin_hist_dist}.  
Gas accretion typically does not cause a systematic shift, but rather smears the initial distribution of $|a|$ (dashed line in panel~(b) of Fig.~\ref{fig:spin_hist_dist}). 
It typically increases the spin along the orbital angular momentum which is frequently reoriented in random directions by binary-single interactions).
In contrast, $|a|$ evolves to $\sim 0.7$ after mergers (solid and dashed black and blue lines). 
Models~M1 and M7 show that $|a_0|$ has an influence on the $|{\bm a}|$ and the $\chi_\mathrm{eff}$ distribution.  
This is seen as the difference between the black and the cyan lines in panels~(a) and (b) in Figs.~\ref{fig:spin_dist}~and~\ref{fig:spin_hist_dist}. 
This suggests that the initial spin distribution might be constrained by GW observations if they originate in AGN disks.

To examine the effects of the initial spin directions, in model~M8 we set $|a_0|=0.7$ and draw random $\hat{\bm a}_0$ directions and in model~M9 we set ${\bm a}_0=0.7 \hat{\bm J}_\mathrm{AGN}$ for all (i.e., including pre-existing) BHs. The resulting distributions for model~M8 are similar to those in model~M6 (orange and brown lines in the top row). This is because the difference between models~M6 and M8 is only the initial spin directions for in-situ formed BHs, whose contribution to all mergers is small ($\sim 1\%$). The alignment of ${\bm a}_0$ with $\hat{\bm J}_\mathrm{AGN}$ in model M9 might be realized if 
previous AGN episodes yield
spin directions aligned with the present-day disk orientation. We include this extreme model as an academic exercise to investigate 
the impact of full initial spin alignment.
In this model, M9, $\theta_a$ is distributed around low values due to the initial direction of $\bm a_0$, while M6 and M8 have broader distributions. The orange and green lines in panel~(a) suggest that the $\bm a_0$ direction has a 
negligible 
influence on the $\chi_\mathrm{eff}$ distribution.

Models~M10--M14 (second row in Figs.~\ref{fig:spin_dist}~and~\ref{fig:spin_hist_dist}; panels~(f)-(j)) show the results when changing the prescriptions or the parameters affecting the evolution of $\hat{\bm a}$ and $\hat{\bm J}_\mathrm{bin}$ during the accretion episodes. 
When the BH spins align efficiently due to enhanced viscosity accelerating the Bardeen-Petterson effect (model~M11) or when the angular momentum of gas captured by the binaries is random (model~M13), $\theta_a$ is slightly closer to an isotropic distribution (orange and brown lines in panel~(h)). Nevertheless, 
we can see that these changes have a small impact on the $\chi_\mathrm{eff}$ distribution (panel~(f)).
This is simply because 
the timescale of randomization of $\hat{\bm J}_\mathrm{bin}$ due to binary-single interactions is shorter than the timescale of alignment of $\hat{\bm a}$ toward $\hat{\bm J}_\mathrm{bin}$ due to gas accretion. Likewise, these prescriptions have very little impact on the distribution of the other quantities.

In the third rows of Figs.~\ref{fig:spin_dist}~and~\ref{fig:spin_hist_dist}, we examine the impact of AGN disk properties. 
It is not clear whether (or to what extent) radial migration operates due to the complexity of the effects of $N$-body migrators \citep{Broz18}, feedback from BHs \citep[][]{delValle18,Regan19}, and inhomogeneities in the turbulent accretion disk \citep{Laughlin2004,Baruteau10}. 
In model~M15 (shown in black lines), radial migration due to torques from the AGN disk is assumed to be inefficient. 
In this model, $\theta_a$ and $\theta_\mathrm{bin}$ (panels~(m) and (n)) are distributed around lower values compared with those in the fiducial model. This is because binaries cannot migrate to high BH-density regions, where 
disorienting 
binary-single interactions are frequent (Paper~I). 
As a result, 
$\chi_\mathrm{eff}$ is distributed toward higher values (panel~(k)). 
Also, $m_\mathrm{chirp}$ tends to be lower since repeated 
``hierarchical'' mergers, which build up the more massive BHs, 
are less frequent (panel~(o)). 
We note here that except for model~M15, the $m_\mathrm{chirp}$ distributions are very similar in all model variants (panels~(e), (j), (o), and (t)). This is because the $m_\mathrm{chirp}$ distributions are determined primarily by how often repeated mergers occur, and this is not significantly influenced by the parameters we changed, other than the efficiency of migration. On the other hand, the $\chi_\mathrm{eff}$ and $m_\mathrm{chirp}$ distributions for a model without migration would be sensitive to various other parameters, such as the initial mass distribution of BHs, and the AGN lifetime. 

In the bottom row of Figs.~\ref{fig:spin_dist}~and~\ref{fig:spin_hist_dist}, in models~M16--M24, 
we examine the influence of the parameters of the AGN disk or the initial BH distribution (see Table \ref{table_results}). 
For example, in Model 18, we reduced the maximum radius at which BHs initially exist $r_\mathrm{out,BH}$, which is roughly equivalent to changing the size of the AGN disk. 
The resulting $\chi_\mathrm{eff}$ distributions are similar in these models (panels~(k) and (p)). 
This is again because the $\chi_\mathrm{eff}$ distribution is mainly affected by how frequently spin-disorienting binary-single interactions take place, which is not sensitive to the changes mentioned above.

Overall, we find that the $\chi_\mathrm{eff}$ distribution is relatively
sensitive to the values of $|a_0|$ and the efficiency of migration, and the $m_\mathrm{chirp}$ distribution is significantly influenced by migration. 
In the next section, we compare these predictions with observations.

\subsection{Comparison with observed distribution}

In Figs.~\ref{fig:spin_obs} and \ref{fig:spin_obs_ligo}, we compare the $\chi_\mathrm{eff}$ distributions predicted by our models with that inferred from the observed GW events. 
In the observed distribution reported by the LIGO and Virgo collaborations \citep{TheLIGO18} 
the $\chi_\mathrm{eff}$ values are concentrated at low absolute values, near zero (Fig.~\ref{fig:spin_obs_ligo}). 
On the other hand, a few possible additional events have been identified with higher and lower $\chi_\mathrm{eff}$ values (Fig.~\ref{fig:spin_obs}, \citealt{Zackay19b,Zackay19,Venumadhav19}; but see also \citealt{Huang20}). 
Note that the IAS group \citep[e.g.][]{Zackay19} 
and the Hannover group \citep{Nitz20} also recovered the events reported by the LIGO/Virgo collaborations.

In Fig.~\ref{fig:mc_xeff_dist}, we additionally show the detection rate distributions predicted in several models in the $\chi_\mathrm{eff}$ vs. $m_\mathrm{chirp}$ plane, together with the distribution inferred from the observed GW events (Table~\ref{table_parameter_data}).

To compare the predicted and the observed distributions quantitatively, we use the KS test as well as a Bayesian analysis. 
The KS test enables us to estimate the probability that the distribution of all or a subset of the observed events is reproduced by a given model, while the Bayesian analysis can be used to assess how consistent each individual event is with a given model.

\subsubsection{Kolmogorov-Smirnov test}
\label{sec:ks_test}

Table~\ref{table_results} lists the results of the KS test.  In each model, 
$P_{\mathrm{KS},\chi_\mathrm{eff}}$, $P_{\mathrm{KS},m_\mathrm{chirp}}$, and 
$P_{\mathrm{KS},\chi_\mathrm{eff},m_\mathrm{chirp}}$ are 
the probabilities that the set of all measured $\chi_\mathrm{eff}$ and $m_\mathrm{chirp}$ values were drawn from the one-dimensional $\chi_\mathrm{eff}$, $m_\mathrm{chirp}$, and the joint two-dimensional 
$(\chi_\mathrm{eff},m_\mathrm{chirp})$--distributions predicted in that model, computed following~\citet{Press88}. 
$P_{\mathrm{KS},\mathrm{LV},\chi_\mathrm{eff}}$ and $P_{\mathrm{KS},\mathrm{LV},m_\mathrm{chirp}}$ are the probabilities that $\chi_\mathrm{eff}$ and $m_\mathrm{chirp}$ values for the events reported by the LIGO/Virgo collaborations were drawn from the predicted one-dimensional $\chi_\mathrm{eff}$ and $m_\mathrm{chirp}$--distributions, respectively. 
Fig.~\ref{fig:mc_xeff_dist} also lists 
$P_{\mathrm{KS},\chi_\mathrm{eff},m_\mathrm{chirp}}$ in each panel. 
As described in Figs.~\ref{fig:spin_dist} and \ref{fig:spin_hist_dist}, 
each predicted merger is weighted by the detectable volume, 
and errors are added on the predicted $\chi_\mathrm{eff}$ and $m_\mathrm{chirp}$ values.

The values of $P_{\mathrm{KS},\chi_\mathrm{eff},m_\mathrm{chirp}}$ are typically $\sim 0.01$--$0.1$, except for model~M15, which yields a much lower value of $P_{\mathrm{KS},\chi_\mathrm{eff},m_\mathrm{chirp}}=6.5\times 10^{-6}$. 
This is because $m_\mathrm{chirp}$ is typically much lower in model~M15 compared 
to the other models, as well as
compared to the observations (panel~(k) in Fig.~\ref{fig:mc_xeff_dist}, $P_{\mathrm{KS},m_\mathrm{chirp}}=9.2\times 10^{-9}$). 
As explained above, this is because in this model (M15), radial migration is turned off; this makes hierarchical mergers much less common.
We note, however, 
that the $m_\mathrm{chirp}$ distribution
has large uncertainties
in our models, for several reasons. First, we do not take into account the exchange of binary components during binary-single interactions, which affects the $m_\mathrm{chirp}$ distribution (the main assumptions in our models are listed in $\S$~2 of Paper~I). Also, the time evolution of the AGN disk model, which we ignore, may affect the merged mass distribution ($\S$~5.7.1 of Paper~I). 
Indeed, using a 30 Myr AGN lifetime in Paper I we found that the mass distribution of merging BHs extends to the values matching the observations if radial migration is turned off  (see Figure 14 therein, panels a and d).
Additional uncertainties include the stellar IMF in galactic centers \citep{Lu13} and  the relation between the initial stellar mass and
its remnant
BH mass \citep[e.g.][]{Belczynski10,Chen15}. 
Particularly, we neglected the possibility that BHs may be delivered to the nuclear star cluster by the infall of low metallicity globular clusters where the BH masses are expected to be higher 
\citep{Tremaine1975,Antonini2013,ArcaSedda18,ArcaSedda19,ArcaSedda20}.
For a more rigorous comparison with the observed $m_\mathrm{chirp}$ distribution, these points should be considered in a future study. 

Focusing only on the $\chi_\mathrm{eff}$ distribution, we find
$P_{\mathrm{KS},\chi_\mathrm{eff}}$ are $\sim 0.1-0.7$. 
This suggests that the observed $\chi_\mathrm{eff}$ distribution 
is consistent with most of our models. 
Fig.~\ref{fig:px_a0} shows $P_{\mathrm{KS},\chi_\mathrm{eff}}$ (black lines) and $P_{\mathrm{KS},\mathrm{LV},\chi_\mathrm{eff}}$ (orange lines) as a function of $|a_0|$ (models~M1--M7). 
The solid and dashed lines show the results in which observational errors are and are not included, respectively. 
$P_{\mathrm{KS},\chi_\mathrm{eff}}$ and $P_{\mathrm{KS},\mathrm{LV},\chi_\mathrm{eff}}$ 
are highest ($0.68$ and $0.15$) in model~M3 and M4 in which $|a_0|=0.2$ and $|a_0|=0.3$, respectively. 
Thus, moderate values for $|a_0|$ are preferred by 
the observed $\chi_\mathrm{eff}$ distribution.

We further investigate how the KS probabilities change if we assume that only some fraction $f_\mathrm{AGN}<1$ of mergers occur in AGN disks, with the remaining fraction $(1-f_\mathrm{AGN})$ produced in other unrelated channel(s).  
This gives an estimate for the maximum allowed fraction of events related to AGN disks in each of our models.
For non-AGN mergers, we 
conservatively 
assume that the $\chi_\mathrm{eff}$ and $m_\mathrm{chirp}$ distributions are the same as the observed distributions to date (Table~\ref{table_parameter_data}), 
but the detection rates are normalized to $(1-f_\mathrm{AGN})$. 
On the other hand, for AGN mergers, 
each merger is weighted by the detectable volume referring to \citet{Kissel18} as before, and 
the total detection rate distribution of $\chi_\mathrm{eff}$ and $m_\mathrm{chirp}$ in each model is normalized to $f_\mathrm{AGN}$.

We construct the $\chi_\mathrm{eff}$ and/or $m_\mathrm{chirp}$ distributions by summing non-AGN and AGN mergers, and calculating the KS probability of their combined distribution. 
The cyan, orange, and black lines in Fig.~\ref{fig:cont_ks} show 
$P_{\mathrm{KS},\chi_\mathrm{eff}}$, $P_{\mathrm{KS},m_\mathrm{chirp}}$, and 
$P_{\mathrm{KS},\chi_\mathrm{eff},m_\mathrm{chirp}}$ 
as a function of $f_\mathrm{AGN}$.
The thick solid lines are the probabilities including all events (Table~\ref{table_parameter_data}), and the dashed lines include only the events reported by the LIGO/Virgo groups \citep{TheLIGO18}. 
Even for model~M15, in which $P_{\mathrm{KS},\chi_\mathrm{eff},m_\mathrm{chirp}}$ is lowest for $f_\mathrm{AGN}=1$, we find
$P_{\mathrm{KS},\chi_\mathrm{eff},m_\mathrm{chirp}}$ is $\gtrsim 0.7$ provided that $f_\mathrm{AGN} \lesssim 0.30$ (solid black line in the right panel). 
Thus, at least $\sim 30\%$ of mergers might originate in AGN disks even in the worst model (see discussion above on caveats which may increase $f_{\rm AGN}$ for model M15).

\subsubsection{Bayesian analysis}
\label{sec:bayesian}

Next, to assess the 
relative
likelihood to produce each event in different models, we calculate the 
Bayes factors
between pairs of models, 
\begin{align}K_{\mathrm{A,B},i}=\frac{P({d}_i|A)}{P({d}_i|B)}
\label{eq:bayesratio}
\end{align}
where $P({d}_i|A)$ is the likelihood of obtaining a data $d_i$ in an event $i$ from Model~$A$, 
\begin{align} P({d}_i|A)=&\nonumber\\\int P({d}_i|m_\mathrm{chirp},\chi_\mathrm{eff},q)&P(m_\mathrm{chirp},\chi_\mathrm{eff},q|A)\nonumber\\&dm_\mathrm{chirp}d\chi_\mathrm{eff}dq\end{align}
where $P({d}_i|m_\mathrm{chirp},\chi_\mathrm{eff},q)$ is the three dimensional likelihood for $m_\mathrm{chirp}$, $\chi_\mathrm{eff}$, and $q$, and $P(m_\mathrm{chirp},\chi_\mathrm{eff},q|A)$ is the probability distribution of $m_\mathrm{chirp}$, $\chi_\mathrm{eff}$, and $q$ in Model~$A$.

To calculate $P(m_\mathrm{chirp},\chi_\mathrm{eff},q|A)$, we first count mergers in $30\times 30 \times 30$ uniform bins in $\chi_\mathrm{eff}$, $m_\mathrm{chirp}$, and $q$ for Model~$A$. 
The maximum and minimum values of $m_\mathrm{chirp}$ for the bins are set to $150$ and $5\,\Msun$, respectively. 
In this procedure, we weighted each merger by the detectable volume. To reduce the statistical fluctuation in the distribution of $\chi_\mathrm{eff}$, $m_\mathrm{chirp}$, and $q$ due to the finite number of mergers in our models, we perform a kernel-density estimate for the distribution using Gaussian kernels whose bandwidth is chosen to satisfy Scott's Rule \citep{Scott92}.

For simplicity, we assume that they follow 
independent Gaussian distributions, as commonly assumed in studies analyzing observed GW data \citep[e.g.][]{Fishbach17}:
\begin{align}&P({d_i}|m_\mathrm{chirp},\chi_\mathrm{eff},q)\nonumber\\    \simeq &N(m_{\mathrm{chirp},i},{\sigma_{m_\mathrm{chirp},i}}^2)N(\chi_{\mathrm{eff},i},{\sigma_{\chi_\mathrm{eff},i}}^2)N(q_{i},{\sigma_{q,i}}^2), \end{align}
where $N(c_1,{c_2}^2)$ is the Gaussian distribution with average $c_1$ and dispersion ${c_2}^2$, $m_{\mathrm{chirp},i}$, $\chi_{\mathrm{eff},i}$ and $q_{i}$ are the median values
and $\sigma_{m_\mathrm{chirp},i}^2$, $\sigma_{\chi_\mathrm{eff},i}^2$, and $\sigma_{q,i}^2$ are the dispersions of $m_{\mathrm{chirp}}$, $\chi_\mathrm{eff}$, and $q$ observed in GW event~$i$. 
We set the average values and the standard deviation for each event to the median values and 
the $90\%$ credible intervals in Table~\ref{table_parameter_data} divided by 3.3,
as appropriate for a Gaussian distribution. 
For the events found by the IAS group, for simplicity, we calculate the dispersion of the source-frame chirp mass assuming no covariance between the parameters.

We calculate $P({d_i}|m_\mathrm{chirp},\chi_\mathrm{eff},q)$ by generating 1000 samples according to the Gaussian distribution and normalizing the distribution as 
\begin{equation}\frac{\int P({d_i}|m_\mathrm{chirp},\chi_\mathrm{eff},q)dm_\mathrm{chirp}d\chi_\mathrm{eff}dq}{\int dm_\mathrm{chirp}d\chi_\mathrm{eff}dq}=1. \end{equation}
The values for $P({d_i}|m_\mathrm{chirp},\chi_\mathrm{eff},q)$ are stored in $30\times 30\times 30$ uniform bins. 

For each event~$i$, we calculate the Bayes factor for a Model~$A$ relative to the observed distribution (i.e. ``Model~B'' in the ratio in Eq.~\ref{eq:bayesratio} is taken to be the observed distribution itself). 
The observed distribution is constructed by smoothing the observed $m_{\mathrm{chirp},i}$, $\chi_{\mathrm{eff},i}$ and $q_i$ distribution using a kernel density estimate as applied above (panel~(t) of Fig.~\ref{fig:mc_xeff_dist}). 
This $K_{\mathrm{A},\mathrm{obs},i}$ presents the strongest test of Model~$A$ for each event, since models are compared
to the actual observed distribution.

In most models, the lowest value of $K_{\mathrm{A},\mathrm{obs},i}$ among the GW sources is typically $\sim 0.02$--$0.2$ ($\mathrm{min}_i K_{\mathrm{A},\mathrm{obs},i}$ in Table~\ref{table_results}), 
except for 
M15, in which it is much lower ($\sim 10^{-5}$). 
In model~M15, $K_{\mathrm{A},\mathrm{obs},i}$ is lowest for GW170817A, which is the source with the highest $m_\mathrm{chirp}$. These findings are consistent with Fig.~\ref{fig:mc_xeff_dist}, which shows that the observed value for the event is outside the predicted range. 
For the fiducial model~M1, we list the value of $K_{\mathrm{A},\mathrm{obs},i}$ for each observed source in Table~\ref{table_parameter_data}. 
The most constraining event is GW151216 ($K_{\mathrm{A},\mathrm{obs},i}=0.050$); the source which has the highest $\chi_\mathrm{eff}$  (see also \citealt{Huang20}).
Thus, as expected, the events with the highest $\chi_\mathrm{eff}$ and $m_\mathrm{chirp}$ constrain the models most strongly.

\subsection{Comparison to other formation channels}
\label{sec:difference}

In this section, we briefly discuss differences in the expected distributions of $\chi_\mathrm{eff}$ and/or $m_\mathrm{chirp}$ between the AGN disk-assisted channel and other binary merger channels.

First, our models with low $|a_0|$ produce the positive correlation between $m_\mathrm{chirp}$ and the dispersion of $\chi_\mathrm{eff}$ 
in $m_\mathrm{chirp}\lesssim 20\,\Msun$ 
(orange lines in Figs.~\ref{fig:mc_xeff_fid} and~\ref{fig:mc_xeff_dist_disp}). 
\citet{Safarzadeh20} estimated that the events reported by the LIGO/Virgo collaborations \citep{TheLIGO18} prefer a positive correlation with $80\%$ confidence. 
Such correlation is somewhat more significant if the events reported by the IAS group are included 
(cyan circles 
in Fig.~\ref{fig:mc_xeff_dist}). 
The field binary evolution channels likely favor rather negative correlation between $m_\mathrm{chirp}$ and the dispersion of $\chi_\mathrm{eff}$ \citep{Gerosa18,Bavera19,Safarzadeh20}. 
The positive correlation is expected for repeated mergers, which frequently occur for multi-body systems in high escape-velocity environments such as galactic nuclei \citep{ArcaSedda20b}, and/or if initial BH spins are low \citep{OLeary16}. 
On the other hand, \citet{ArcaSedda20b} show that the positive correlation is not reproduced by mergers in dynamical environments. 
Hence, the positive correlation suggested by \citet{Safarzadeh20} may be a signature that the observed mergers are facilitated in AGN disks.

Second, AGN disks can produce high--$m_\mathrm{chirp}$ mergers. 
For field binaries, $m_\mathrm{chirp}$ is limited to $\lesssim 40\,\Msun$ due to  pair instability supernovae \citep[e.g.][]{Kinugawa14,Spera19}. In the scenarios involving dynamical formation and evolution, \citet{ArcaSedda20b} predicted that $99\%$ of mergers have $m_\mathrm{chirp}\lesssim 50\,\Msun$. For mergers in AGN disks, we find that $\sim 10$--$15\%$ of mergers have $m_\mathrm{chirp}\gtrsim 50\,\Msun$ if BHs migrate efficiently. 
Thus, if mergers with $m_\mathrm{chirp}\gtrsim 50\,\Msun$ are discovered, they would favor the AGN-disk origin.
Although a false alarm rate is high (0.34 $\mathrm{yr}^{-1}$), \cite{Udall19} reported a high-mass binary BH merger event, GW170502, with a chirp mass of $\sim 70\,\Msun$. Similar events with high S/N ratio will be an additional signature for mergers in AGN disks.

Third, if migration is inefficient (model~M15), $\chi_\mathrm{eff}$ can be negative and the $\chi_\mathrm{eff}$ distribution may lack symmetry around $\chi_\mathrm{eff}=0$. From Table~\ref{table_results}, the absolute value for the 90 percentiles for $\chi_\mathrm{eff}$ ($|\chi_\mathrm{eff,90}|$) is larger than that for the 10 percentiles ($|\chi_\mathrm{eff,10}|$) by $\sim 0.27$ in this model. Such asymmetric distribution of $\chi_\mathrm{eff}$ is caused by gas accretion, while it is reduced by the randomization of the binary angular momentum directions due to binary-single interactions.

Mergers in isolated environments are unlikely to produce negative $\chi_\mathrm{eff}$ \citep{Bavera19}, unless angular momentum transfer is inefficient; however in this case high--$\chi_\mathrm{eff}$ mergers are overproduced \citep{Belczynski17c}. 
Mergers in globular clusters and galactic nuclei (without AGN disks) can produce negative $\chi_\mathrm{eff}$, but the $\chi_\mathrm{eff}$ distribution is almost perfectly symmetric around $\chi_\mathrm{eff}=0$~\citep{Rodriguez18PRL}. 
Hence, if the asymmetric distribution is observed, a possible interpretation is that mergers originate in AGN disks and binary-single interactions are less efficient (e.g. due to inefficient inward migration to the densely populated inner regions).

In summary, the positive correlation between $m_\mathrm{bin}$ and the dispersion of $\chi_\mathrm{eff}$, 
high--$m_\mathrm{chirp}$ mergers, 
and an asymmetric $\chi_\mathrm{eff}$ distribution might be possible signatures that distinguish mergers in AGN disks from other channels. However, the $\chi_\mathrm{eff}$ and $m_\mathrm{chirp}$ distributions for mergers in AGN disks are found to be strongly affected by radial migration of BHs.  The efficiency of this migration is still poorly understood, and should be investigated in the future. 

\subsection{Consistency with GW190412}

\label{sec:gw190412}

Recently, a low mass-ratio event, GW190412, has been reported \citep{LIGO20_GW190412}. 
This is the first event which has a low mass ratio ($q=0.28^{+0.13}_{-0.07}$), is constrained to have non-zero BH spin parallel to the binary's orbital plane ($\chi_p = 0.30^{+0.19}_{-0.15}$), 
has a primary BH with large spin ($a_1=0.43^{+0.16}_{-0.26}$), 
and is chosen from $\sim 50$ third observing run triggers. 
\citet{Fishbach20} predicted that $99\%$ of mergers have $q>0.51$ from the LIGO/Virgo events in O1/O2, which suggests that GW190412 is a highly unusual event. 
\citet{Gerosa20} and \citet{SafarzadehHotokezaka2020} have shown that this event is exceedingly rare in both scenarios for mergers in isolated fields and globular clusters due to its low mass ratio and high projected spin component. 
Here, we suggest that the properties of GW190412 can be naturally explained by higher-generation mergers in an AGN disk.  Due to the low mass ratio, we can expect that this may be a merger between a first-generation secondary BH with $M_2\approx 8~{\rm M_\odot}$ and a second- (or higher-) generation primary BH with $M_1\approx 30~{\rm M_\odot}$. Indeed, $q\sim 0.28$ and $m_\mathrm{bin}\sim 38\,\Msun$ is common for mergers in AGN disks (see Fig.~14 in Paper~I). 
Furthermore, since mergers endow the remnant BH with high spin, the high value for the primary BH spin in GW190412 is consistent with it having experience one or more prior mergers.    Finally, in our models, the AGN disk delivers BHs to the inner regions where binary-single interactions frequently misalign the spins relative to the orbital angular momentum.
If we include this event to the analysis in $\S$~\ref{sec:bayesian}, the Bayes factor between model~M1 and the observed distribution for GW190412 is 0.70, which suggests that the AGN channel can naturally explain the properties of GW190412 well.

\section{Conclusions}

In this paper we investigated the distribution of the effective spin parameter $\chi_\mathrm{eff}$ for BH binaries merging in accretion disks of AGN.  We performed one-dimensional $N$-body simulations, combined with semi-analytical prescriptions of the relevant processes.   
$\chi_\mathrm{eff}$ is enhanced by the alignment of BH spins toward the binary orbital angular momenta due to gas accretion, while it is reduced by the randomization of binary orbital angular momenta due to hard binary-single interactions. 
This is the first detailed estimate for the $\chi_\mathrm{eff}$ distribution of stellar--mass BH mergers in AGN disks, considering the effects of binary-single interactions and gas accretion. 
Our main results can be summarized as follows:

\begin{enumerate}

\item 
Due to the randomization of the binary orbital angular momentum directions by frequent binary-single interactions, 
$\chi_\mathrm{eff}$ is symmetric around zero, 
if radial migration of BHs to the inner, densely populated regions is efficient. 
The median value of $|\chi_\mathrm{eff}|$ depends most strongly 
on the initial BH spin magnitudes and the efficiency of migration, and is much less impacted by the other parameters or prescriptions we considered. 

\item 
The $\chi_\mathrm{eff}$ distribution for all observed events 
reported by the LIGO/Virgo collaborations and the IAS group 
during LIGO/Virgo O1 and O2 is roughly consistent with the distribution expected for mergers in AGN disks (Fig.~\ref{fig:spin_obs}). 
The KS probabilities between the $\chi_\mathrm{eff}$ distribution of all events and those in our models are typically $\sim 0.1-0.7$ ($P_{\mathrm{KS},\chi_\mathrm{eff}}$, Table~\ref{table_results}). The observed $\chi_\mathrm{eff}$ distribution 
slightly favors
moderate values for the initial BH spins ($|a_0|\lesssim 0.5$; see Fig.~\ref{fig:px_a0}).

\item 
Even for the worst-fitting model, the fractional contribution of mergers in AGN disks to all observed mergers is limited only to $\lesssim 0.3$ (Fig.~\ref{fig:cont_ks}), and much higher contributions are allowed in our other models.

\item 
The positive correlation between $m_\mathrm{chirp}$ and the dispersion of $\chi_\mathrm{eff}$ can be reproduced by AGN-assisted mergers if the initial BH spin magnitude is low (Figs.~\ref{fig:mc_xeff_fid} and~\ref{fig:mc_xeff_dist_disp}, $\S$~\ref{sec:difference}). 
Also, mergers in AGN disks might be distinguished from other channels based on
the chirp masses extending to values as high as $\approx 300~{\rm M_\odot}$ (see also Paper I).

\item 
The properties of the recently announced gravitational-wave event, GW190412, including  a low mass ratio, a high spin for the primary BH, and a spin component in the orbital plane, are naturally expected if it is a hierarchical merger in an AGN disk.

\end{enumerate}

\begin{table*}
	\caption{
		The results in different models. 
		The first two columns show the model number and indicate its variation from the fiducial model (M1).
        For example, in M16 (``No gas hard'') binaries are
        not hardened by gas interaction, and M15 (``w/o gas mig.'') excludes type I/II torques and the resulting radial migration in the AGN disk.  
        In the next four columns,  
        $\chi_\mathrm{eff,med}$, $\chi_\mathrm{eff,10}$,   $\chi_\mathrm{eff,90}$ are the median, 10 percentile, and 90 percentile for the $\chi_\mathrm{eff}$-- and
         the median for the $|\chi_\mathrm{eff}|$--distributions, respectively, 
       in which observational errors are included.  
       In the next three columns,
       $P_{\mathrm{KS},\chi_\mathrm{eff}}$, $P_{\mathrm{KS},m_\mathrm{chirp}}$, and  $P_{\mathrm{KS},\chi_\mathrm{eff},m_\mathrm{chirp}}$ are, respectively, the KS probabilities that the all observed events were drawn from 
       the $\chi_\mathrm{eff}$--, $m_\mathrm{chirp}$--, and the joint ($\chi_\mathrm{eff},m_\mathrm{chirp})$--distributions predicted in each model. 
       In the next two columns, $P_{\mathrm{KS},\mathrm{LV},\chi_\mathrm{eff}}$ and  $P_{\mathrm{KS},\mathrm{LV},m_\mathrm{chirp}}$ are 
       the KS probabilities that the events reported by the LIGO/Virgo collaborations were drawn from the predicted $\chi_\mathrm{eff}$-- and $m_\mathrm{chirp}$--distributions. 
       In the last column, $\mathrm{min}_i K_{\mathrm{A},\mathrm{obs},i}$ is 
       the lowest value 
       of the Bayes factor among all observed GW events,  evaluated for each event between the given model and the observed distribution itself. 
        	}
\label{table_results}
\hspace{-18mm}
\begin{tabular}{c|c||c|c|c|c|c|c|c|c|c|c}
\hline
\multicolumn{2}{c}{input} \vline& \multicolumn{10}{c}{output}\\\hline
Model&Parameter&
${\chi}_\mathrm{eff,med}$&
${\chi}_\mathrm{eff,10}$&
${\chi}_\mathrm{eff,90}$&
$|{\chi}_\mathrm{eff}|_\mathrm{med}$&
$P_{\mathrm{KS},\chi_\mathrm{eff}}$&
$P_{\mathrm{KS},m_\mathrm{chirp}}$&
$P_{\mathrm{KS},\chi_\mathrm{eff},m_\mathrm{chirp}}$
&$P_{\mathrm{KS},\mathrm{LV},\chi_\mathrm{eff}}$
&$P_{\mathrm{KS},\mathrm{LV},m_\mathrm{chirp}}$
&$\mathrm{min}_i K_{\mathrm{A},\mathrm{obs},i}$

\\\hline

M1&Fiducial&
0.012&-0.29&0.33
&0.15&
0.53&0.25&0.12
&0.10&0.35
&0.050
\\\hline

M2&$a_0=0.1$&
0.022&-0.31&0.36
&0.17&
0.44&0.18&0.091
&0.088&0.25
&0.052
\\\hline

M3&$a_0=0.2$&
0.022&-0.35&0.37&0.19&
0.68&0.15&0.069
&0.14&0.23
&0.078
\\\hline

M4&$a_0=0.3$&
0.020&-0.37&0.43&0.21&
0.59&0.14&0.033
&0.15&0.18
&0.12
\\\hline

M5&$a_0=0.5$&
0.032&-0.43&0.47&0.26&
0.54&0.060&0.022
&0.12&0.12
&0.15
\\\hline

M6&$a_0=0.7$&
0.043&-0.47&0.55&0.29&
0.41&0.18&0.036
&0.093&0.28
&0.21
\\\hline

M7&$a_0=0.99$&
0.022&-0.57&0.63
&0.34&
0.31&0.19&0.030
&0.061&0.26
&0.18
\\\hline

M8&random $\hat{\bm a}_0$, ${a}_0=0.7$&
0.036&-0.48&0.54
&0.29&
0.45&0.17&0.030
&0.091&0.25
&0.21
\\\hline

M9&${\bm a}_0=0.7 \hat{\bm z}$&
0.018&-0.56&0.63
&0.35&
0.23&0.095&0.082
&0.049&0.55
&0.19
\\\hline

M10&$\nu_2/\nu_1=2$&
0.021&-0.30&0.35
&0.16&
0.46&0.086&0.052
&0.10&0.15
&0.087
\\\hline

M11&$\nu_2/\nu_1=50$&
0.017&-0.29&0.32
&0.15&
0.56&0.11&0.045
&0.11&0.17
&0.039
\\\hline

M12&$\hat{{\bm J}}_\mathrm{CBHD}=\hat{\bm z}$&
0.012&-0.33&0.35
&0.18&
0.45&0.20&0.097
&0.081&0.29
&0.045
\\\hline

M13&$f_\mathrm{rot}=0$&
0.0026&-0.29&0.31
&0.14&
0.61&0.26&0.14
&0.12&0.41
&0.068
\\\hline

M14&$f_\mathrm{rot}=10$&
0.017&-0.31&0.32
&0.16&
0.62&0.24&0.092
&0.14&0.28
&0.075
\\\hline

M15&w/o gas mig.&
0.12&-0.19&0.46
&0.19&
0.077&$9.2\times 10^{-9}$&$6.5\times 10^{-6}$
&0.068&$2.1\times 10^{-4}$
&$9.8\times 10^{-6}$
\\\hline

M16&No gas hard.&
0.014&-0.32&0.35
&0.17&
0.75&0.55&0.036
&0.19&0.11
&0.054
\\\hline

M17&${\dot M}_\mathrm{out}={\dot M}_\mathrm{Edd}$&
0.0095&-0.31&0.33&0.16&
0.68&0.15&0.078
&0.15&0.22
&0.064
\\\hline

M18&$r_\mathrm{out,BH}=0.3$ pc&
0.0023&-0.29&0.30&0.15&
0.23&0.020&0.021
&0.036&0.057
&0.021
\\\hline

M19&$M_\mathrm{SMBH}=4\times 10^7\Msun$&
0.017&-0.31&0.35&0.17&
0.31&0.065&0.028
&0.066&0.13
&0.11
\\\hline

M20&$\delta_\mathrm{IMF}=-1.7$&
0.032&-0.29&0.35&0.16&
0.54&0.16&0.046
&0.12&0.23
&0.064
\\\hline

M21&${\beta}_\mathrm{v}=1$&
-0.0046&-0.32&0.33&0.16&
0.31&0.0014&0.0038
&0.058&0.011
&0.025
\\\hline

M22&$M_\mathrm{star,3pc}=3\times 10^6\Msun$&
0.016&-0.31&0.34&0.17&
0.28&0.033&0.037
&0.054&0.080
&0.048
\\\hline

M23&$\gamma_{\rho}=1.5$&
0.017&-0.31&0.34&0.16&
0.63&0.16&0.083
&0.14&0.21
&0.058
\\\hline

M24&twice $m_\mathrm{BH}$&
0.041&-0.29&0.35&0.17&
0.60&0.15&0.062
&0.13&0.14
&0.033
\\\hline

M1&$t=3$ Myr&
0.014&-0.23&0.28&0.12&
0.87&0.18&0.26
&0.24&0.43
&0.031
\\\hline

M1&$t=30$ Myr&
0.016&-0.37&0.38&0.19&
0.49&0.082&0.043
&0.11&0.15
&0.096
\\\hline

\end{tabular}
\end{table*}

\begin{table*}
\begin{center}
\caption{The data sets used in this paper, adopted from [1]: \citet{TheLIGO18}, [2]:~\citet{Zackay19}, [3]:~\citet{Venumadhav19}, and [4]:~\citet{Zackay19b}. 
Note that reference [1] quotes the source-frame, whereas [2,3] quote the detector-frame chirp masses, together with their respective errors (columns 2 and 3, respectively). 
For the events found by [2-4], we calculate the dispersion of the source-frame chirp mass assuming no covariance between the parameters. 
$K_{\mathrm{M1,obs},i}$ is the Bayes factor between model~M1 and the observed distribution for each event $i$. }
\label{table_parameter_data}
\hspace{-5mm}
\begin{tabular}{c|c|c|c|c|c|c|c}
\hline 
Event  & 
$M_\mathrm{chirp}$& 
$M_\mathrm{chirp}^\mathrm{det}$&
$\chi_\mathrm{eff}$&$q$&$z$&Reference
&$K_{\mathrm{M1,obs},i}$
\\
\hline\hline
GW150914&$28.6_{-1.5}^{+1.6}$&
-&
$-0.01_{-0.13}^{+0.12}$&$0.87_{-0.21}^{+0.12}$&$0.09_{-0.03}^{+0.03}$&[1]
&0.29
\\\hline

GW151012&$15.2_{-1.1}^{+2.0}$&
-&
$0.04_{-0.19}^{+0.28}$&$0.59_{-0.34}^{+0.36}$&$0.21_{-0.09}^{+0.09}$&[1]
&0.77
\\\hline

GW151226&$8.9_{-0.3}^{+0.3}$&
-&
$0.18_{-0.12}^{+0.20}$&$0.56_{-0.33}^{+0.38}$&$0.09_{-0.04}^{+0.04}$&[1]
&0.88
\\\hline

GW170104&$21.5_{-1.7}^{+2.1}$&
-&
$-0.04_{-0.20}^{+0.17}$&$0.65_{-0.22}^{+0.30}$&$0.19_{-0.08}^{+0.07}$&[1]
&0.51
\\\hline

GW170608&$7.9_{-0.2}^{+0.2}$&
-&
$0.03_{-0.07}^{+0.19}$&$0.70_{-0.36}^{+0.27}$&$0.07_{-0.02}^{+0.02}$&[1]
&1.4
\\\hline

GW170729&$35.7_{-4.7}^{+6.5}$&
-&
$0.36_{-0.25}^{+0.21}$&$0.68_{-0.28}^{+0.28}$&$0.48_{-0.20}^{+0.19}$&[1]
&0.23
\\\hline

GW170809&$25.0_{-1.6}^{+2.1}$&
-&
$0.07_{-0.16}^{+0.16}$&$0.67_{-0.23}^{+0.29}$&$0.20_{-0.07}^{+0.05}$&[1]
&0.36
\\\hline

GW170814&$24.2_{-1.1}^{+1.4}$&
-&
$0.07_{-0.11}^{+0.12}$&$0.83_{-0.23}^{+0.15}$&$0.12_{-0.04}^{+0.03}$&[1]
&0.44
\\\hline

GW170818&$26.7_{-1.7}^{+2.1}$&
-&
$-0.09_{-0.21}^{+0.18}$&$0.76_{-0.24}^{+0.21}$&$0.20_{-0.07}^{+0.07}$&[1]
&0.30
\\\hline

GW170823&$29.3_{-3.2}^{+4.2}$&
-&
$0.08_{-0.22}^{+0.20}$&$0.76_{-0.28}^{+0.22}$&$0.34_{-0.14}^{+0.13}$&[1]
&0.28
\\\hline

GW151216&
$22\pm3$&
$31_{-3}^{+2}$&
$0.81_{-0.21}^{+0.15}$&$0.7_{-0.3}^{+0.3}$&$0.43_{-0.17}^{+0.17}$&[2]
&0.050
\\\hline

GW170121&
$23\pm4$&
$29_{-3}^{+4}$&
$0.3_{-0.3}^{+0.3}$&$0.76_{-0.26}^{+0.19}$&$0.24_{-0.13}^{+0.14}$&[3]
&0.39
\\\hline

GW170304&
$31\pm7$&
$47_{-7}^{+8}$&

$0.2_{-0.3}^{+0.3}$&$0.75_{-0.25}^{+0.19}$&$0.5_{-0.2}^{+0.2}$&[3]
&0.27
\\\hline

GW170727&
$29\pm 6$&
$42_{-6}^{+6}$&
$-0.1_{-0.3}^{+0.3}$&$0.7_{-0.3}^{+0.2}$&$0.43_{-0.17}^{+0.18}$&[3]
&0.30
\\\hline

GW170425&
$31 \pm 15$&
$47_{-10}^{+26}$&
$0.0_{-0.5}^{+0.4}$&$0.6_{-0.3}^{+0.3}$&$0.5_{-0.3}^{+0.4}$&[3]
&0.44
\\\hline

GW170202&
$17\pm 3$&
$21.6_{-1.4}^{+4.2}$&
$0.2_{-0.3}^{+0.4}$&$0.5_{-0.2}^{+0.4}$&$0.27_{-0.12}^{+0.13}$&[3]
&0.72
\\\hline

GW170403&
$33\pm 7$&
$48_{-7}^{+9}$&
$-0.7_{-0.3}^{+0.5}$&$0.7_{-0.3}^{+0.2}$&$0.45_{-0.19}^{+0.22}$&[3]
&0.22
\\\hline

GW170817A&$41\pm 7$&
-&
$0.5_{-0.2}^{+0.2}$&$0.7$&$0.6_{-0.2}^{+0.2}$&[4]
&0.17
\\\hline

\end{tabular}
\end{center}
\end{table*}

\begin{figure*}\begin{center}\includegraphics[width=180mm]{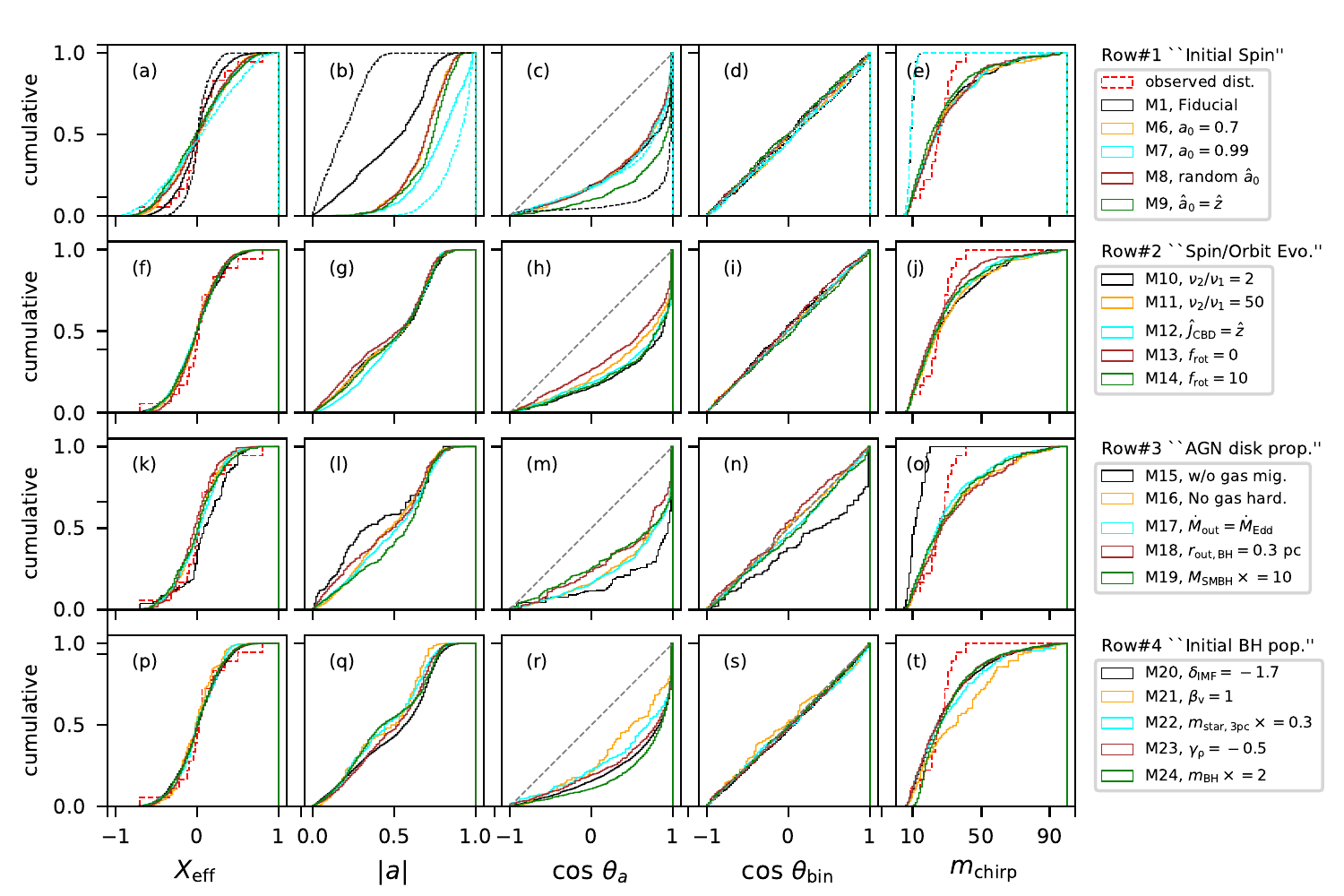}\caption{Same as Fig.~\ref{fig:spin_hist_dist}, but show distributions without errors. }\label{fig:spin_hist_ne}\end{center}\end{figure*}

\begin{figure*}\begin{center}
\includegraphics[width=195mm]{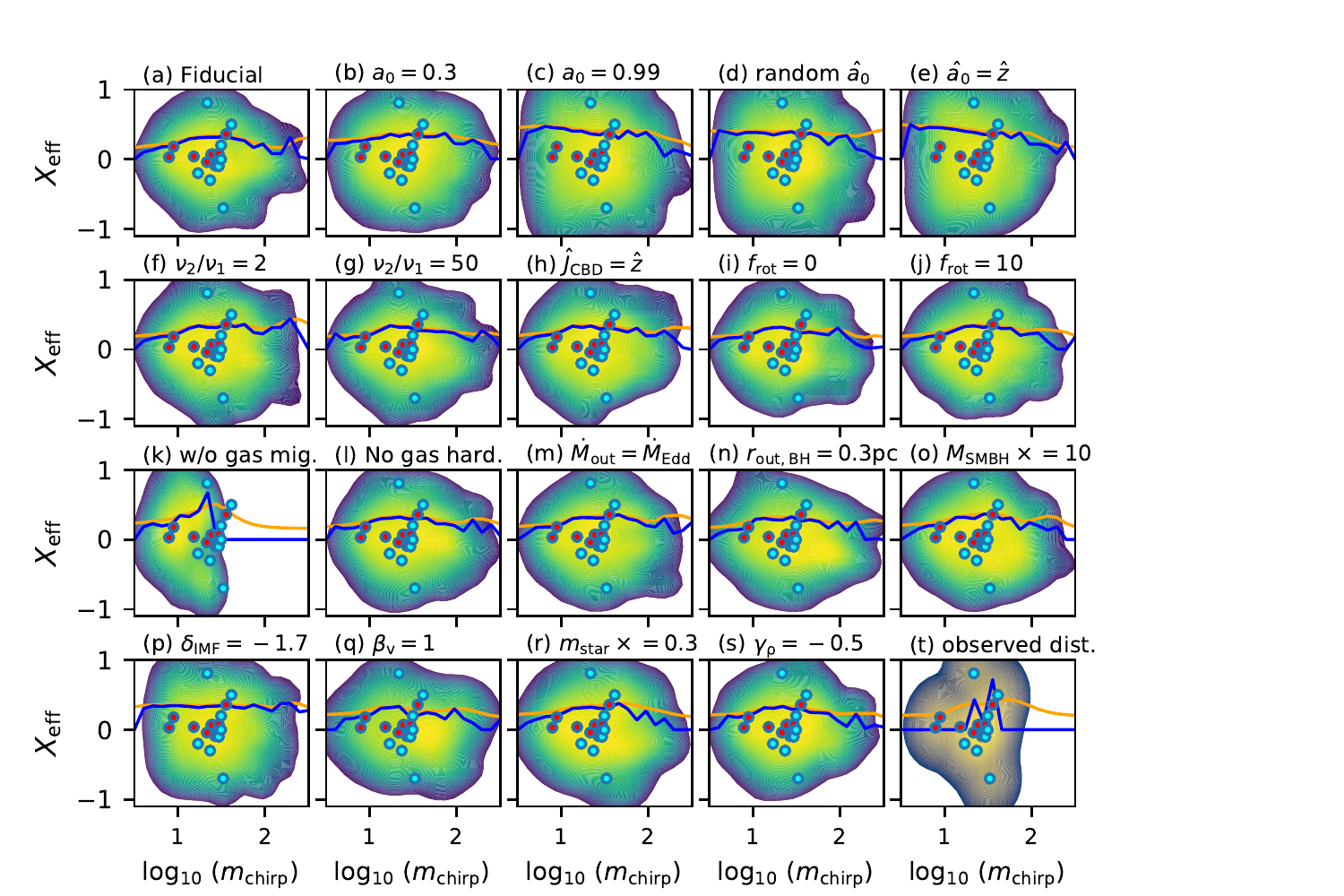}
\caption{
Same as Fig.~\ref{fig:mc_xeff_dist}, but the standard deviations are presented. 
Blue lines show the standard deviation of the mergers in each bin, and orange lines show the standard deviation of the $\chi_\mathrm{eff}$ distribution which is produced by performing a kernel-density estimate. The latter is shown as an estimate of the trend for the observed distribution in which the number of events is small.
}\label{fig:mc_xeff_dist_disp}
\end{center}\end{figure*}

\acknowledgments

We thank Barry McKernan, Brian Metzger, Saavik Ford, Leigh Nathan for useful discussions.
This project has received funding from the European Research Council (ERC) under the European Union's Horizon 2020 research and innovation programme under grant agreement No 638435 (GalNUC) and by the Hungarian National Research, Development, and Innovation Office grant NKFIH KH-125675. 
 ZH acknowledges support from NASA grant NNX15AB19G and NSF grant 1715661.
Simulations and analyses were carried out on Cray XC50 and computers 
at the Center for Computational Astrophysics, National Astronomical Observatory of Japan.

\bibliographystyle{yahapj.bst}
\bibliography{agn_bhm}

\end{document}